# Stacking-Dependent Interlayer Phonons in 3R and 2H MoS$_2$

Jeremiah van Baren[1⊥], Gaihua Ye[2⊥], Jia-An Yan[3], Zhipeng Ye[2], Pouyan Rezaie[2], Peng Yu[4], Zheng Liu[5], Rui He[2*], Chun Hung Lui[1*]

[1] *Department of Physics and Astronomy, University of California, Riverside, California 92521, United States*

[2] *Department of Electrical and Computer Engineering, Texas Tech University, Lubbock, Texas 79409, United States*

[3] *Department of Physics, Astronomy, and Geosciences, Towson University, Towson, Maryland 21252, United States*

[4] *School of Materials Science and Engineering, Sun Yat-sen University, Guangzhou 510275, Guangdong, China*

[5] *School of Materials Science and Engineering, Nanyang Technological University, Singapore 637371, Singapore*

⊥ *These authors contributed equally*
\* *Corresponding authors:* rui.he@ttu.edu, joshua.lui@ucr.edu

**Abstract:** We have investigated the interlayer shear and breathing phonon modes in MoS$_2$ with pure 3R and 2H stacking order by using polarization-dependent ultralow-frequency Raman spectroscopy. We observe up to three shear branches and four breathing branches in MoS$_2$ with thickness from 2 to 13 layers. The breathing modes show the same Raman activity behavior for both polytypes, but the 2H breathing frequencies are consistently several wavenumbers higher than the 3R breathing frequencies, signifying that 2H MoS$_2$ has slightly stronger interlayer lattice coupling than 3R MoS$_2$. In contrast, the shear-mode Raman spectra are strikingly different for 2H and 3R MoS$_2$. While the strongest shear mode corresponds to the highest-frequency branch in the 2H structure, it corresponds to the lowest-frequency branch in the 3R structure. Such distinct and complementary Raman spectra of the 3R and 2H polytypes allow us to survey a broad range of shear modes in MoS$_2$, from the highest to lowest branch. By combining the linear chain model, group theory, effective bond polarizability model and first-principles calculations, we can account for all the major observations in our experiment.

## 1. Introduction

Molybdenum disulfide (MoS$_2$) is the most commonly studied two-dimensional (2D) transition metal dichalcogenide (TMD). Monolayer MoS$_2$ exhibits a S-Mo-S trigonal prismatic (1H) structure with a lack of inversion symmetry (Figure 1). The non-centrosymmetric crystalline structure, combined with 2D quantum confinement and strong spin-orbit coupling, produce many remarkable properties in monolayer MoS$_2$, including the direct band gap [1, 2], spin-valley coupling [3], optical harmonic generation [4, 5], magnetoelectricity [6], and piezoelectricity [7]. Beyond monolayer, MoS$_2$ commonly exhibits the hexagonal 2H stacking order, in which the adjacent layers are rotated by 180$^o$ and stack directly upon one another (Figure 1). For even numbers of layers, 2H MoS$_2$ restores the inversion symmetry and hence loses many interesting valleytronic, piezoelectric and nonlinear optical properties [4, 5, 7-9]. This greatly limits the applications of MoS$_2$. Fortunately, MoS$_2$ hosts another stable polytype with the rhombohedral 3R stacking order, in which the adjacent layers displace slightly from each other without any rotation [10, 11] (Figure 1). This distinct polytype can break the inversion symmetry, and hence retain excellent valleytronic, piezoelectric and nonlinear optical properties, for all layer numbers [10-14]. In addition, 3R MoS$_2$ shows superior catalytic properties to 2H MoS$_2$ [15] and can possibly host distinctive excitonic states [16]. These recent developments have stimulated much scientific interest in 3R MoS$_2$.

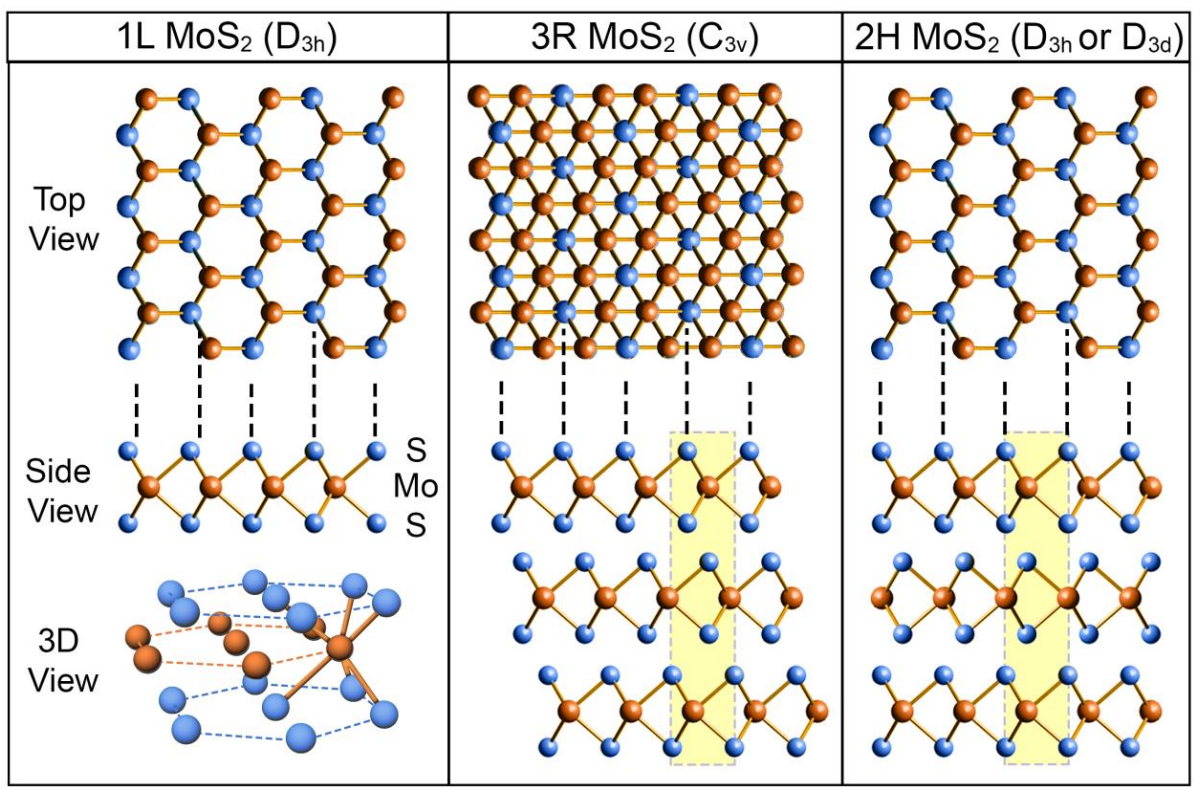

**Figure 1.** The crystalline structure of monolayer MoS$_2$, few-layer 3R and 2H MoS$_2$ polytypes. The top (bottom) column shows the top (side) view of the crystals. The dashed boxes highlight the interlayer atomic alignment. 3R MoS$_2$ has the non-centrosymmetric point group C$_{3v}$ for all layer numbers. 2H MoS$_2$ has the D$_{3h}$ (D$_{3d}$) point group for odd (even) layer numbers.

Although 3R MoS$_2$ atomic crystals exhibit attractive properties, there is a lack of efficient and reliable methods to characterize their layer number and stacking order. Conventional methods such as transmission electron microscopy, *x*-ray diffraction, and scanning probe measurements cannot determine both the layer number and stacking order accurately for multi-layer crystals; they are also difficult to apply to microscale samples. In this respect, Raman spectroscopy of interlayer phonons can provide rapid and nondestructive identification of the layer number and stacking sequence. 2D materials generally exhibit two types of interlayer phonon modes, namely the shear (S) and breathing (B) modes, which arise from interlayer vibrations with tangential and vertical layer displacement, respectively. As they are generated entirely by interlayer coupling, the interlayer phonons are highly sensitive to the layer thickness [17], stacking order [18], and surface environment [19]. Therefore, ultralow-frequency Raman spectroscopy of interlayer phonons has been widely applied to characterize the interlayer coupling and stacking structure in many 2D materials, including graphene [17-26], TMDs [27-61], boron nitride [62, 63], and black phosphorene [64-68]. Researchers have just recently studied the interlayer Raman modes in 3R TMD atomic crystals [56-61]; however, the results are ambiguous because they used either natural crystals or samples grown by chemical vapor deposition (CVD), both of which contain random mixture of different stacking sequences. To this point, a comprehensive investigation of interlayer phonons in pure 3R MoS$_2$ atomic crystals is still lacking.

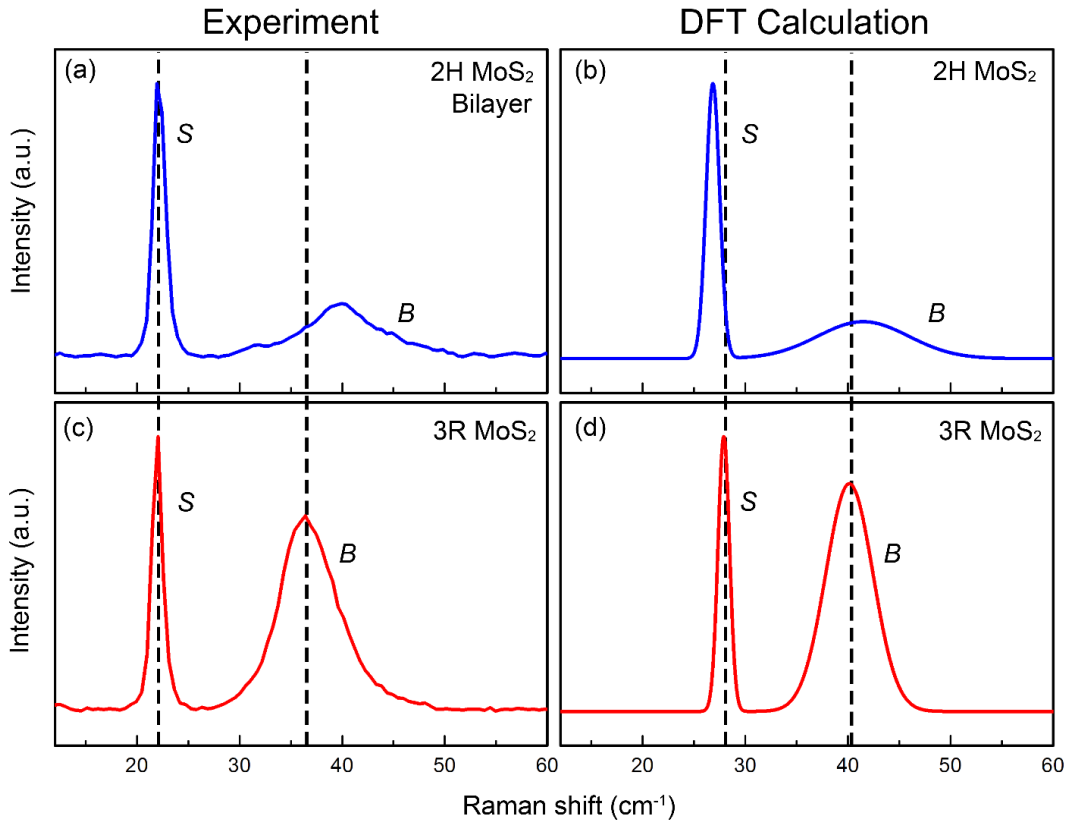

**Figure 2.** Measured (a, c) and calculated (b, d) Raman spectra of the interlayer shear (S) and breathing (B) modes for 2H (a, b) and 3R (c, d) MoS$_2$ bilayer. The theoretical spectra are obtained by first-principle calculations based on density functional theory (DFT); they are presented as Gaussian functions of the same line width as the experimental spectra. The dashed lines highlight the different frequency separation between the shear and breathing modes in the 2H and 3R polytypes.

Here, we report a comparative Raman study of interlayer phonon modes in pure 3R and 2H $MoS_2$ atomic crystals with thickness from 2 to 13 layers. We observe up to four breathing branches and three shear branches in the low-frequency Raman spectra. The interlayer shear modes exhibit distinct layer-number dependence and branch dependence between the 3R and 2H polytypes. While 2H $MoS_2$ exhibits the strongest Raman signal in the highest-frequency shear branch, 3R $MoS_2$ exhibits the strongest Raman response in the lowest-frequency shear branch. Such opposite Raman behavior reflects the different crystal symmetry in the 3R and 2H phases. On the other hand, the breathing modes exhibit similar layer-number and branch dependence for both polytypes, but the 2H breathing modes are consistently higher in frequency (~3 $cm^{-1}$) than the 3R breathing modes. This result indicates a small but detectable difference in the interlayer coupling strength between the 3R and 2H stacking order. We can explain all major observations by a combined analysis of the linear chain model, group theory, effective bond polarizability model and first-principles calculations.

## 2. Experimental methods

We fabricated 2H and 3R $MoS_2$ samples with layer number $N = 1$ to 13 (labeled as 1L – 13L) by mechanical exfoliation of bulk crystals onto $Si/SiO_2$ substrates. We used natural 2H $MoS_2$ crystals from HQ Graphene Inc, and grew 3R $MoS_2$ single crystals by the chemical vapor transport (CVT) method [11]. In the synthesis of 3R $MoS_2$ crystals, we used $MoCl_5$ as the transport agent. The stoichiometric amount of $Mo:S:MoCl_5$ is 9:20:1 and the total mass is 450 mg. They were sealed in an evacuated 20-cm-long quartz tube under vacuum of $10^{-6}$ Torr. The tube was placed in a three-zone furnace. The reaction zone was pretreated at 850 °C for 30 hours with the grown zone at 900 °C. The reaction zone was then programed at a higher temperature 1060 °C with the growth zone at a lower temperature 920°C for six days to provide a temperature gradient for the crystal growth. Finally, the furnace was naturally cooled down to room temperature and the 3R $MoS_2$ single crystals were collected in the growth zone. Our crystals show uniform 3R stacking order with almost no mixture of other phases. Such high-quality 3R $MoS_2$ crystals enable us to exfoliate atomically thin layers with pure 3R stacking order.

We measure the Raman spectra of $MoS_2$ samples using a commercial Horiba LabRam HR Raman microscope in the back-scattering geometry at room temperature. The excitation light source is a 532 nm continuous laser. The incident laser power on the samples was below 1 mW, with a spot diameter of ~1 μm. We measure both unpolarized and polarized Raman spectra. In the former case, we collect Raman signals of all polarizations. In the latter case, we excite the samples with either horizontally (H) or vertically (V) polarized laser and collect the Raman signals at only the vertical (V) polarization. Correspondingly, we obtain the parallel-polarized (VV) and cross-polarized (HV) Raman spectra. The shear modes in $MoS_2$, being doubly-degenerate, have two-dimensional group representations with finite diagonal and off-diagonal elements in their Raman tensors; they are therefore observed in both the VV and HV geometries. But the breathing modes, being non-degenerate, have one-dimensional group representations with only diagonal elements in the Raman tensors; they only appear in the VV geometry.

## 3. Experimental Results

We first present the unpolarized Raman spectra of 3R and 2H MoS$_2$ bilayers, the simplest pair to compare the two polytypes (Figure 2a, c). Both polytypes exhibit a shear mode and a breathing mode, but their spectra show two different features. First, while the frequencies of their shear modes are nearly the same (22 cm$^{-1}$), the 3R breathing mode (37 cm$^{-1}$) is lower in frequency than the 2H breathing mode (40 cm$^{-1}$). This implies that the 3R layers are less strongly bound than the 2H layers. Second, the intensity ratio between the breathing and shear modes is much higher in 3R MoS$_2$ than in 2H MoS$_2$. The breathing-mode peak intensity is ~70% (~25%) of the shear-mode intensity for the 3R (2H) MoS$_2$ bilayers.

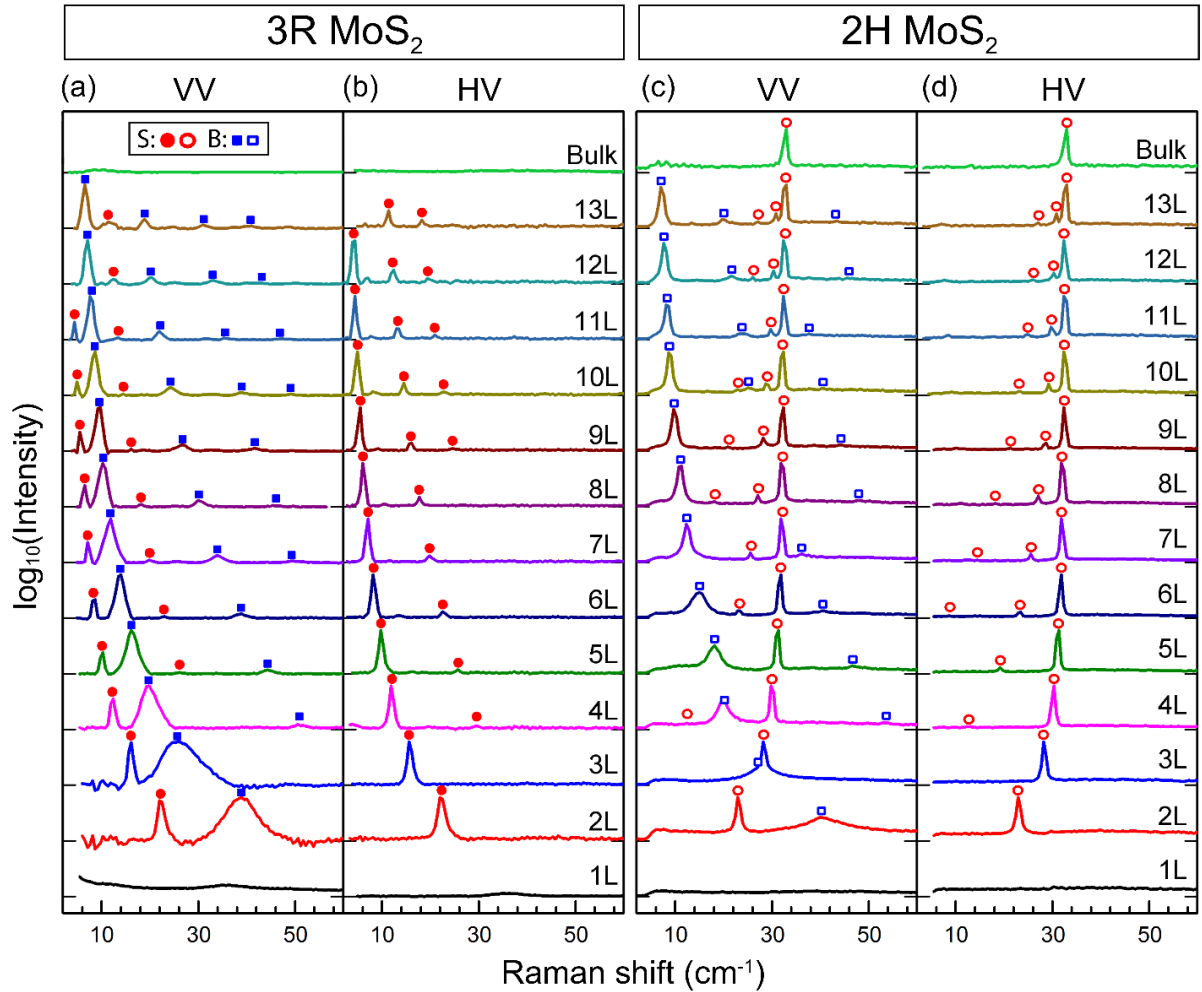

**Figure 3.** (a-d) parallel-polarized (VV) and cross-polarized (HV) Raman spectra of the shear modes (S; red dots) and breathing modes (B; blue squares) for 3R MoS$_2$ (solid symbols) and 2H MoS$_2$ (open symbols) with layer number $N = 1 - 13$. The breathing modes appear only in the VV geometry, whereas the shear modes appear in both the VV and HV geometries. We have subtracted the broad Raman background and plot the spectra in the logarithmic scale to display the weak interlayer modes.

We have calculated the unpolarized Raman spectra of the shear and breathing modes for both 3R and 2H MoS$_2$ bilayers by density functional theory (DFT) (Figure 2b, d; see Supplementary Material). Our calculations reproduce the two spectral features discussed above. First, the separation between the shear and breathing modes is 3 cm$^{-1}$ smaller in 3R MoS$_2$ than in 2H MoS$_2$ in our simulated spectra. This result matches the observed 3 cm$^{-1}$ red shift of the 3R breathing mode relative to the 2H breathing mode. Second, the intensity ratio between the calculated breathing and shear modes agrees well with the experimental ratio. Our experiment and calculation therefore consistently show that 3R and 2H MoS$_2$ have noticeable difference in the interlayer coupling strength and Raman response. We can use the above two major spectral features to distinguish the two types of MoS$_2$ bilayers.

Thicker MoS$_2$ samples display more interlayer Raman modes. Figure 3 displays the logarithmic parallel-polarized (VV) and cross-polarized (HV) Raman spectra for both 3R and 2H MoS$_2$ with layer numbers $N = 1 - 13$. We have subtracted the broad Raman background to display the weak interlayer modes (see Supplementary Material for details). We observe up to three branches of shear modes and four branches of breathing modes in thick MoS$_2$ samples. The 3R and 2H breathing modes have similar thickness dependence. For both polytypes, the strongest breathing Raman peak redshifts as the layer number increases (blue squares in Figure 3a, c). In contrast, the 3R and 2H shear modes show strikingly different spectra as the layer thickness increases. Figure 3b, d display the shear-mode spectra in the HV geometry, in which all breathing modes are suppressed. For the 2H polytype, the shear Raman peaks blueshift as the layer number increases. But for the 3R polytype, the shear Raman peaks redshift as the layer number increases. Therefore, the 2H and 3R shear Raman modes evolve oppositely with the layer number.

## 4. Theoretical discussion and analysis

### 4.1. Linear chain model

We can describe the frequencies of the breathing (shear) modes by a simple linear chain model in which adjacent layers are coupled harmonically with the same force constant [17]. An $N$-layer system hosts $N$-1 breathing modes and $N$-1 doubly-degenerate shear modes. Their mode frequencies are:

$$\omega_{N,B}^{(j)} = \omega_B \cos\left(\frac{j\pi}{2N}\right) \qquad (1)$$

$$\omega_{N,S}^{(j)} = \omega_S \cos\left(\frac{j\pi}{2N}\right) \qquad (2)$$

Here $j = 1, 2, \ldots N$-1 is the branch index, counting from the highest-frequency to the lowest-frequency branch; $\omega_B$ ($\omega_S$) is the frequency of the highest breathing (shear) branch in the bulk limit. These two bulk frequencies are the only fitting parameters in the linear chain model.

Figure 4 compares the observed breathing and shear mode frequencies (symbols) with the prediction of the linear chain model (lines). The observed breathing modes in 3R and 2H MoS$_2$ can be fit well with a bulk breathing-mode frequency of $\omega_B =$ 53.85 cm$^{-1}$ and 57.40 cm$^{-1}$, respectively. The strongest breathing Raman mode corresponds to the lowest breathing branch ($j = N - 1$), and other weaker breathing Raman modes correspond to higher branches with $j = N - 3$, $N - 5$, $N - 7$ … (solid blue lines in Figure 4). Other breathing branches with $j = N - 2$, $N - 4$, $N - 6$ … are Raman inactive (dashed blue lines in Figure 4).

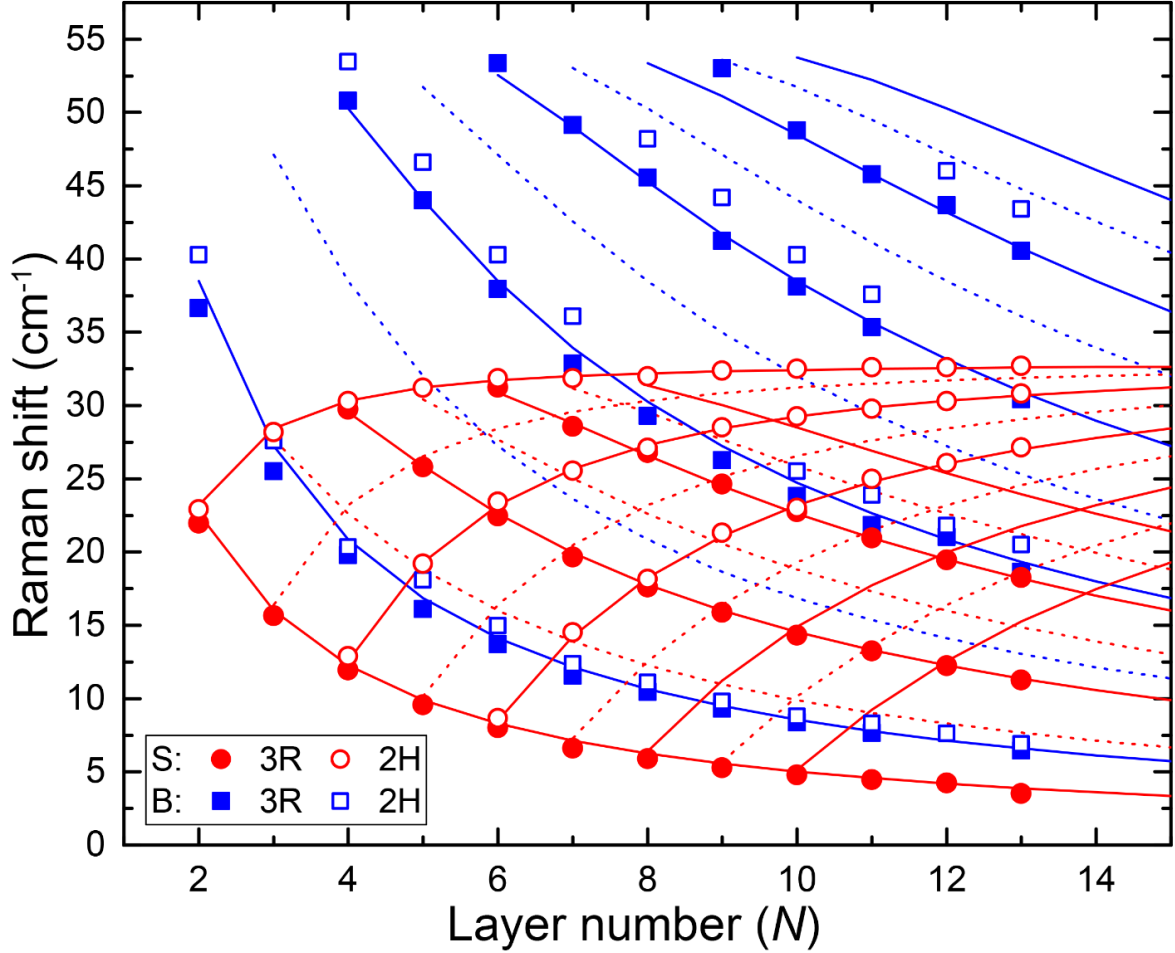

**Figure 4.** Frequencies of the observed shear modes (S; red) and breathing modes (B; blue) in Figure 3 as a function of MoS$_2$ layer number. The solid (open) symbols are experimental data for 3R (2H) MoS$_2$. The blue (red) lines are the predicted frequencies of the breathing (shear) modes by the linear chain model, as described in the text. The solid (dashed) lines denote the Raman active (inactive) branches. The 2H breathing modes are consistently higher in frequency than the 3R breathing modes, whereas the shear modes have almost the same frequencies for both polytypes.

For the shear modes, the observed Raman frequencies in 3R and 2H MoS$_2$ can be fit well with a bulk frequency $\omega_S =$ 31.85 cm$^{-1}$ and 32.85 cm$^{-1}$, respectively. Both values match closely the measured shear frequency (32.9 cm$^{-1}$) of bulk 2H MoS$_2$ (the top spectrum in Figure 3c-d). The linear chain model shows that the observed shear modes in 2H MoS$_2$ correspond to high-frequency shear branches with $j = 1, 3, 5$ (red open dots and solid lines in Figure 4). The branches with $j =$ 2, 4, 6 … are Raman inactive (red dashed lines). In contrast, the observed shear modes in 3R MoS$_2$ correspond to low-frequency shear branches with $j = N - 1, N - 3, N - 5$ (red solid dots). The branches with $j = N - 2, N - 4, N - 6$ … are Raman inactive (red dashed lines). The observed 2H and 3R shear modes therefore cover the high and low shear branches, respectively. For MoS$_2$ with odd layer numbers, the observable 2H and 3R shear modes complement each other. By combining the 2H and 3R data, we can reveal all the shear branches ($j = 1$ to $N - 1$) in MoS$_2$ with odd layer numbers $N = 3, 5, 7$.

### 4.2. Interlayer force constants and elastic moduli

Our measured interlayer mode frequencies allow us to calculate the interlayer force constants and elastic moduli of 3R and 2H MoS$_2$ (Table 1). The interlayer force constant (K) is related to the bulk interlayer vibration frequency ($\omega$) as $K = (\omega \pi c)^2 \mu$, where $c$ is the speed of light and $\mu = 3.068 \times 10^{-6}$ kg m$^{-2}$ is the mass per unit area of MoS$_2$ monolayer [17, 31]. By using the extracted 3R and 2H bulk shear mode frequencies, we obtain the interlayer shear force constants: $K_x = 2.76 \times 10^{19}$ Nm$^{-3}$ for 3R MoS$_2$ and $K_x = 2.94 \times 10^{19}$ Nm$^{-3}$ ($K_x$ and $K_y$ are the same because MoS$_2$ is isotropic in the $xy$ plane). Similarly, by using the extracted bulk breathing mode frequencies, we obtain the interlayer compressive force constants: $K_z = 7.89 \times 10^{19}$ Nm$^{-3}$ for 3R MoS$_2$ and $K_z = 8.98 \times 10^{19}$ Nm$^{-3}$ for 2H MoS$_2$. The elastic modulus (C) is related to the force constant as $C = Kt$, where $t$ is the interlayer separation. 2H and 3R MoS$_2$ have slightly different interlayer distance: $t = 6.148$ Å for 2H MoS$_2$ and $t = 6.123$ Å for 3R MoS$_2$. Correspondingly, we obtain the shear modulus $C_{44} = K_x t = 16.90$ GPa (18.08 GPa) for 3R (2H) MoS$_2$ and the compressive modulus $C_{33} = K_z t = 48.32$ GPa (55.19 GPa) for 3R (2H) MoS$_2$ (Table 1). Our 2H results agree with prior works [30, 31]. Notably, the 2H compressive modulus is 14% higher than the 3R modulus; this ratio matches roughly the prediction of recent first-principles calculations [69].

| | Elastic parameters | 3R MoS$_2$ | 2H MoS$_2$ |
|---|---|---|---|
| Shear Properties | $\omega_S$ (cm$^{-1}$) | 31.85 | 32.85 |
| | $K_x$ (10$^{19}$ Nm$^{-3}$) | 2.76 | 2.94 |
| | $C_{44}$ (GPa) | 16.90 | 18.08 |
| Compressive Properties | $\omega_B$ (cm$^{-1}$) | 53.85 | 57.40 |
| | $K_z$ (10$^{19}$ Nm$^{-3}$) | 7.89 | 8.98 |
| | $C_{33}$ (GPa) | 48.32 | 55.19 |

**Table 1.** Elastic parameters of 2H and 3R MoS$_2$ calculated from the interlayer phonon frequencies

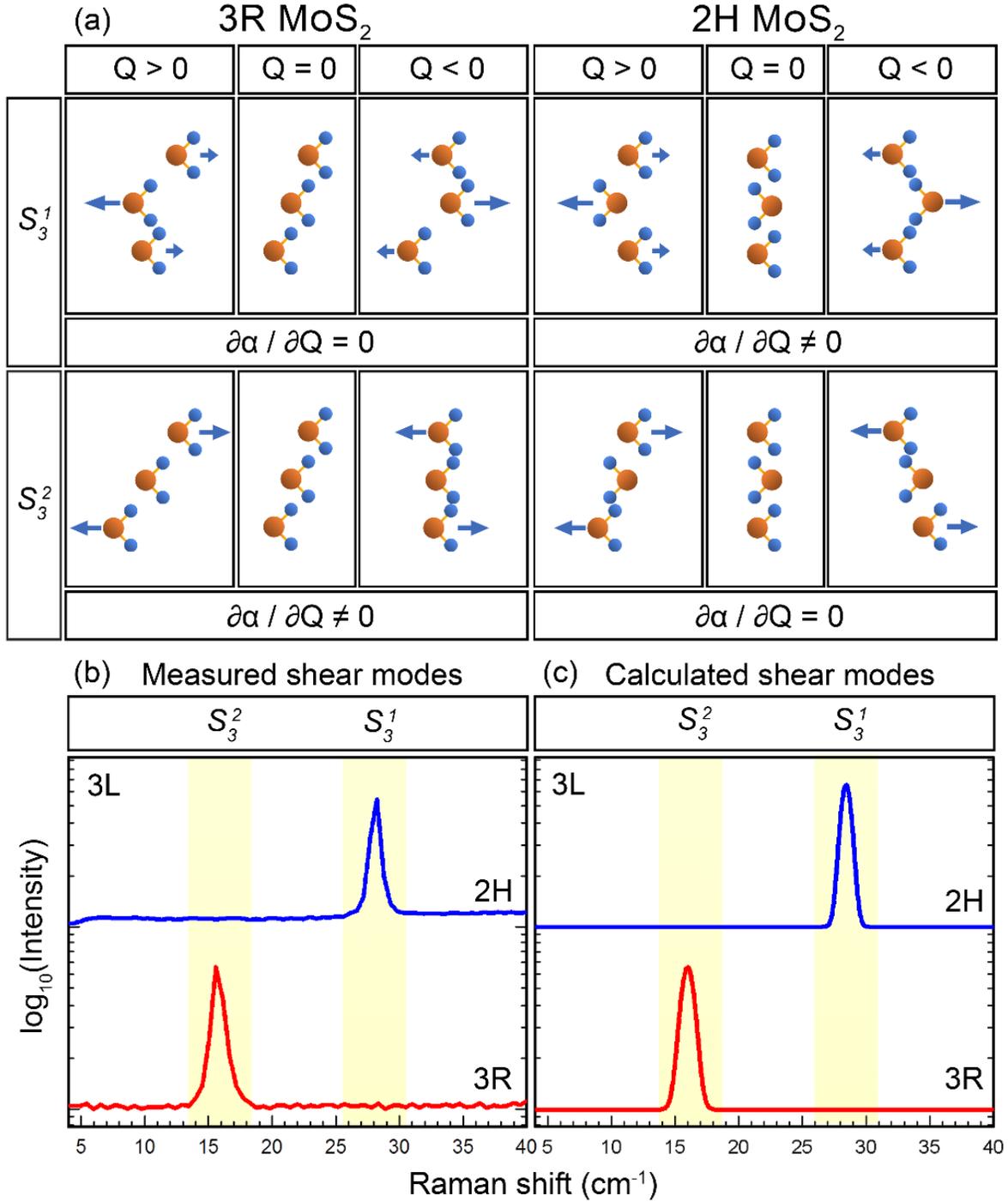

**Figure 5.** (a) Layer displacement of the two shear modes in 3R and 2H MoS$_2$ trilayers. By comparing the atomic configuration at opposite normal displacements ($Q > 0$ and $Q < 0$) near the equilibrium position ($Q = 0$), we can deduce whether the differential polarizability ($\partial\alpha/\partial Q$) (hence, the Raman tensor) is zero or finite. (b-c) The experimental and simulated cross-polarized Raman shear modes for trilayer 2H and 3R MoS$_2$. In the simulation, the shear-mode frequency is predicted by the linear chain model (highlighted by the yellow bars) and the relative mode intensity is determined by the bond polarizability model. The high and low shear modes appear exclusively in the 2H and 3R trilayer, respectively.

### 4.3. Group-theory analysis

After calculating the elastic parameters, we examine the Raman properties of interlayer phonons. The interlayer phonon modes in 2H MoS$_2$ have been well studied in the literature [28, 31, 70, 71]. Their Raman activity behavior can be understood from their point group symmetry. 2H MoS$_2$ with even (odd) layer numbers have the $D_{3d}$ ($D_{3h}$) point group. From the representations of the shear (breathing) modes, we can easily deduce that they are Raman active in every other branch, counting from the highest (lowest) branch, while being Raman inactive in other branches (see the Supplementary Material for a summary of group-theory analysis for 2H MoS$_2$). However, the point-group analysis is not applicable in the 3R structure, which has lower symmetry than the 2H structure. 3R MoS$_2$ has the $C_{3v}$ point group for all the layer numbers, with no inversion nor mirror symmetry. All breathing modes have the $A_1$ representation and all shear modes have the $E$ representation. They are all Raman active in the group-theory analysis. Therefore, a simple point-group symmetry analysis cannot explain the observed alternating appearance of the interlayer branches in the 3R Raman spectra.

### 4.4. Qualitative description of Raman activity

To reveal the underlying physics of interlayer Raman modes in the 3R structure, we directly examine the influence of stacking order on the Raman tensor of the phonon modes. For a vibration mode $k$, its Raman intensity ($I_k$) is proportional to [72, 73]:

$$I_k \propto \frac{(n_k+1)}{\omega_k} \left| \hat{e}_i \cdot \tilde{R}_k \cdot \hat{e}_s^T \right|^2 \tag{3}$$

Here $\omega_k$ is the phonon frequency; $n_k$ is the phonon occupation according to the Bose-Einstein distribution; $\hat{e}_i$ and $\hat{e}_s$ are the polarization unit vectors of the incident and scattered light, respectively; $\tilde{R}_k$ is the Raman tensor. In a classical picture, $\tilde{R}_k$ can be expressed as:

$$\tilde{R}_k = \frac{\partial \tilde{\alpha}}{\partial Q_k} \Delta Q_k \tag{4}$$

Here $\tilde{\alpha}$ is the polarizability tensor; $Q_k$ is the normal coordinate of the vibration normal mode; $\partial \tilde{\alpha}/\partial Q_k$ is the derivative of the polarizability tensor at the equilibrium lattice position ($Q_k = 0$).

To determine the Raman activity of a phonon mode, we can directly examine the atomic configuration during the vibration. As an illustration, Figure 5a shows the interlayer atomic configurations of the two shear modes $S_3^1$ and $S_3^2$ in 3L MoS$_2$ at positive ($Q > 0$) and negative ($Q < 0$) normal displacement. The 3R and 2H structures have distinct atomic configurations during the vibration. For the 3R $S_3^1$ mode, the interlayer atomic configurations are the same for $Q > 0$ and $Q < 0$. The differential polarizability $\partial \alpha/\partial Q$ is thus zero; this mode is Raman inactive. But the $S_3^1$ mode in the 2H structure has distinct atomic configurations for $Q > 0$ and $Q < 0$. This mode thus has non-zero $\partial \alpha/\partial Q$ and is Raman active. In contrast, the lower $S_3^2$ mode has opposite stacking dependence. The 3R $S_3^2$ mode exhibits different interlayer atomic configurations for $Q > 0$ and $Q < 0$, leading to finite $\partial \alpha/\partial Q$ and Raman response. However, the 2H $S_3^2$ mode has the same atomic configurations for $Q > 0$ and $Q < 0$, leading to zero $\partial \alpha/\partial Q$ and no Raman response. This simple analysis illustrates why the high and low shear branches appear exclusively in the Raman spectra from 3L MoS$_2$ of 2H and 3R phases, respectively (Figure 5b). The above pictorial analysis can be applied to explain the Raman activity of shear and breathing modes in 2H and 3R structures with any number of layers (e.g. see the analysis of 5L MoS$_2$ in the Supplementary Material). Similar stacking-dependent shear modes are also observed in few-layer graphene with ABA and ABC stacking order [18, 26].

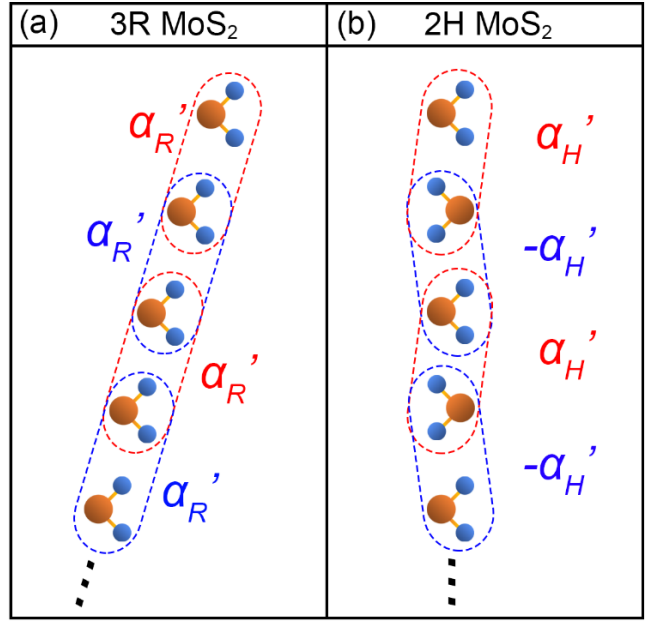

**Figure 6.** Schematic representation of interlayer bonds for (a) 3R and (b) 2H MoS$_2$. The interlayer bonds have the same lateral orientation for 3R stacking, but alternate lateral orientation for 2H stacking, resulting in their distinct shear-mode Raman response. The blue/red dashed lines highlight the interactions between adjacent layers

### 4.5. Quantitative bond polarizability model

Our qualitative analysis above can be quantified in an effective bond polarizability model [72, 73], which predicts the relative Raman intensity of different shear modes. This model considers the distortion of interlayer bonds by the interlayer vibration. The Raman tensor $\tilde{R}$ is the sum of contributions from each interlayer bond. The diagonal element $R_{xx}$ is responsible for parallel-polarized Raman response; the off-diagonal element $R_{xy}$ is responsible for cross-polarized Raman response. For a shear mode $k$ with $x$-direction layer displacement in an $N$-layer structure, the tensor element $R_{\mu\nu}$ ($\mu, \nu = x, y, z$) may be expressed as:

$$R_{\mu\nu}(k) = \sum_{l=1}^{N-1} \alpha'_l (x_l - x_{l+1}). \tag{5}$$

Here $l$ denotes the layer index; $x_l$ is the lateral displacement of layer $l$ from the equilibrium position; $\alpha_l$ is the $\mu\nu$-element of the polarizability tensor of the interlayer bond between layers $l$ and $l+1$. $\alpha'_l = \partial \alpha_l / \partial x_l$ is the derivative of $\alpha_l$ with respect to $x_l$. For a shear mode $S_N^j$ with branch index $j$ in an $N$-layer structure, the displacement of layer $l$ is: [17]

$$x_l \propto \cos\left[\frac{(N-j)(2l-1)}{2N}\pi\right] \tag{6}$$

The 3R and 2H structure have the same layer displacement pattern in vibration, but different configurations of interlayer bonds. In 3R MoS$_2$, all the layers have the same orientation and identical interlayer bonds, and hence the same $\alpha'_l$ (Figure 6a). We may thus remove the $l$ index and name it as $\alpha'_l = \alpha'_R$. By Eq. 5, the 3R Raman tensor element is:

$$R_{\mu\nu}(k) = \alpha'_R[(x_1 - x_2) + (x_2 - x_3) + \cdots + (x_{N-1} - x_N)] = \alpha'_R(x_1 - x_N). \tag{7}$$

It depends only on the displacement of the top and bottom layers.

In contrast, the adjacent layers in 2H MoS$_2$ have opposite orientation. The interlayer bonds thus exhibit alternate lateral orientation for successive layer pairs, leading to opposite signs of $\alpha'_l$ (Figure 6b). By setting $\alpha'_{l=1} = \alpha'_H$, the other $\alpha'_l$ would be $(-1)^{l+1}\alpha'_H$. By Eq. 5, the 2H Raman tensor element is:

$$R_{\mu\nu}(k) = \sum_{l=1}^{N-1}(-1)^{l+1}\alpha'_H(x_l - x_{l+1})$$
$$= \alpha'_H[(x_1 - x_2) - (x_2 - x_3) + (x_3 - x_4) - \cdots (x_{N-1} - x_N)]. \tag{8}$$

We can combine Eq. 3, 6, 7 and 8 to obtain the Raman intensity. After some algebraic derivation (see Supplementary Material), the Raman intensities of the 3R and 2H shear modes $S_N^j$ show the following analytic forms:

$$I_{3R}(k) \propto \frac{n_k+1}{\omega_k}|\alpha'_R|^2 \sin^2\left[\frac{(N-j)(N-1)}{2N}\pi\right]\left[1 - (-1)^{N-j}\right] \tag{9}$$

$$I_{2H}(k) \propto \frac{n_k+1}{\omega_k}|\alpha'_H|^2 \sin^2\left(\frac{N-j}{2N}\pi\right)\tan^2\left(\frac{N-j}{2N}\pi\right)\left[1 - (-1)^j\right] \tag{10}$$

In Eq. 9, the factor $1 - (-1)^{N-j}$ is zero when $N - j$ is an even number; the 3R shear mode is non-zero only when the branch index $j = N$ -1, $N$ -3, $N$-5, …. In Eq. 10, the factor $1 - (-1)^j$ is zero when $j$ is an even number; the 2H shear mode is non-zero only when $j = 1, 3, 5, …$. Such stacking-dependent Raman activity exactly matches our experimental results (Figure 3, 4).

In addition to accounting for the Raman activity, the above analytical theory enables us to quantitatively simulate the Raman spectra of shear modes for each layer number and stacking order. In the simulation, we use the linear chain model (Eq. 1, 2) to determine the mode frequencies and the bond polarizability model (Eq. 9, 10) to determine the Raman intensity ratio of different branches. The calculation gives the same results for parallel and cross polarizations. For comparison with experiment, we plot the theoretical Raman modes as Gaussian functions with the same width as the experimental peaks. Figure 5 and 7 compare our simulations with experiment for 3L, 5L and 9L MoS$_2$, which exhibit one, two and three shear modes, respectively, for both the 3R and 2H phases in our observed Raman spectra. Our calculated spectra show reasonably accurate intensity ratio between different shear branches (see Supplementary Material for simulations for other layer numbers).

Our model can be readily extended to explain the breathing modes by redefining $\alpha'_l = \partial \alpha_l / \partial z_l$, where $z_l$ is the vertical displacement of layer $l$. In this case, $\alpha'_l$ does not depend on the lateral orientation of the interlayer bonds, so $\alpha'_l$ is the same for all layer indices $l$ in either stacking order. As a result, for both 2H and 3R breathing modes, their Raman tensor and intensity follow the same forms as those of 3R shear modes in Eq. 7 and 9. For instance, Figure 8 displays our calculated breathing-mode Raman spectrum for 10L MoS$_2$ (the same for 2H and 3R phases), in comparison with the measured parallel-polarized (VV) spectrum of 10L 3R MoS$_2$. Our calculation quantitatively reproduces the experimental spectrum, which exhibits four breathing modes ($B_{10}^9$, $B_{10}^7$, $B_{10}^5$, $B_{10}^3$) from low to high frequency and strong to weak intensity.

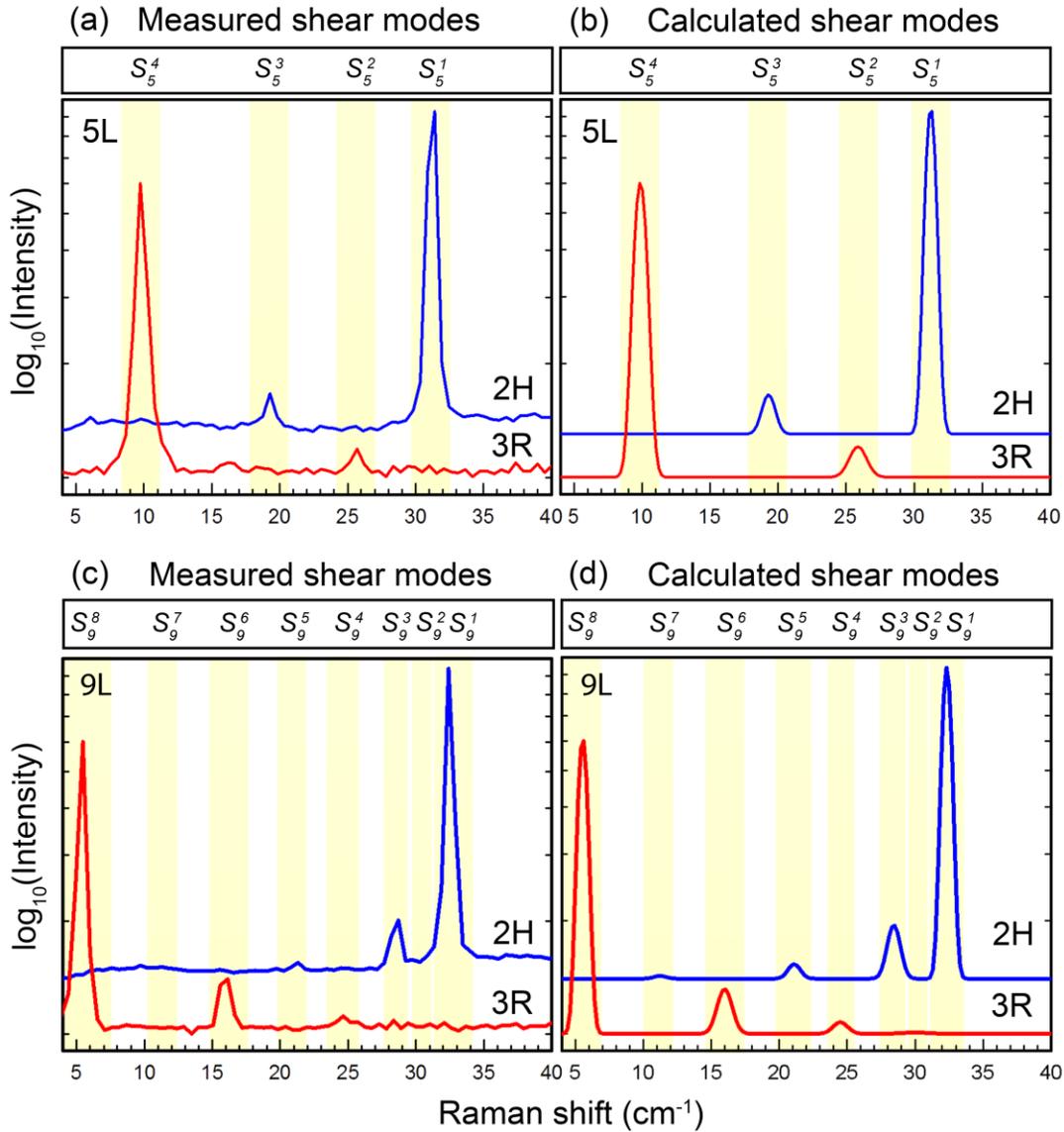

**Figure 7.** Comparison of the cross-polarized (HV) experimental and simulated shear-mode Raman spectra for 5L and 9L MoS$_2$ in both 3R and 2H phases. In the simulation, the shear-mode frequency is predicted by the linear chain model (highlighted by the yellow bars) and the relative mode intensity is determined by the bond polarizability model.

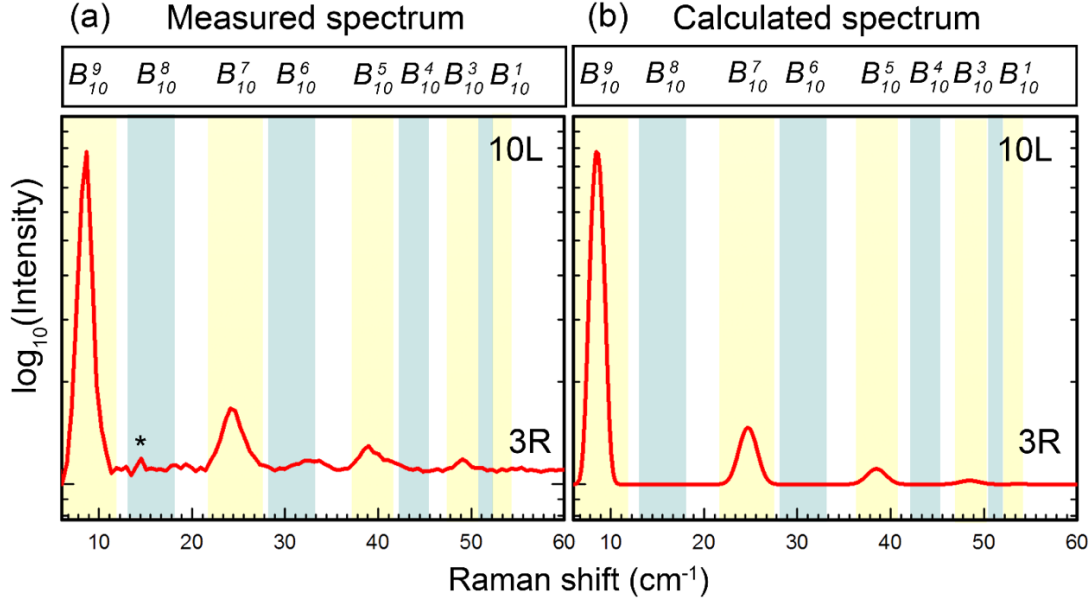

**Figure 8.** (a) The experimental parallel-polarized spectrum of 10L 3R MoS$_2$. The four major peaks are the breathing modes. The asterisk denotes a weak shear mode. (b) The simulated breathing-mode spectrum for 10L MoS$_2$ by the linear chain model and the bond polarizability model. The yellow (blue) bars denote the Raman active (inactive) breathing modes

## 5. Conclusion

We have carried out a comprehensive experimental and theoretical investigation of the interlayer Raman modes in 3R and 2H MoS$_2$ atomic layers. The 3R and 2H polytypes exhibit similar breathing-mode spectra but distinct shear-mode spectra. Prior research has measured the interlayer Raman modes in various 2D materials with different stacking order, such as few-layer graphene with ABA and ABC stacking order [18, 26, 74], TMDs with 2H structure [27-31], TiSe$_2$ with 3R stacking order [51], *etc*. Nonetheless, due to the difficulties in sample preparation and Raman experiment, no prior work could systematically measure both the shear and breathing modes in the same material at two distinct stacking orders and all layer numbers. Our comparative analysis of 3R and 2H MoS$_2$ of layer number $N = 1 - 13$ thus represents a key Raman study to address the role of stacking order and layer number in 2D materials. In addition, our results serve to facilitate the research of 3R TMD materials, whose general lack of inversion symmetry may enable many novel valleytronic, piezoelectric, and nonlinear optical applications.

**Acknowledgement:** Work at TTU (G.Y., Z.Y., R.H.) is supported by NSF CAREER Grant No. DMR-1760668. The DFT calculations of this work used the Extreme Science and Engineering Discovery Environment (XSEDE) Comet at the SDSC through allocation TG-DMR160101 and TG-DMR160088. J.A.Y. acknowledges support from the NSF grant DMR 1709781, the Fisher General Endowment, and SET grants from the Jess and Mildred Fisher College of Science and Mathematics at Towson University. P.Y. and Z.L. acknowledge support from the Singapore National Research Foundation under NRF Award Nos. NRF-RF2013-08.

# Supplementary Material for
# Stacking-Dependent Interlayer Phonons in 3R and 2H MoS$_2$

Jeremiah van Baren[1⊥], Gaihua Ye[2⊥], Jia-An Yan[3], Zhipeng Ye[2], Pouyan Rezaie[2], Peng Yu[4], Zheng Liu[5], Rui He[2*], Chun Hung Lui[1*]

[1] *Department of Physics and Astronomy, University of California, Riverside, California 92521, United States*

[2] *Department of Electrical and Computer Engineering, Texas Tech University, Lubbock, Texas 79409, United States*

[3] *Department of Physics, Astronomy, and Geosciences, Towson University, Towson, Maryland 21252, United States*

[4] *School of Materials Science and Engineering, Sun Yat-sen University, Guangzhou 510275, Guangdong, China*

[5] *School of Materials Science and Engineering, Nanyang Technological University, Singapore 637371, Singapore*

⊥ *These authors contributed equally*
\* *Corresponding authors: rui.he@ttu.edu, joshua.lui@ucr.edu*

# Table of Contents

## 1. Baseline subtraction for the ultralow-frequency Raman spectra

When we measure the Raman spectra of interlayer modes, we observe a broad background in the low-frequency range. The Raman spectra in the main paper are presented after the subtraction of the background. Here we show our procedure to substrate the background signals (Figure S1).

One major background component is a broad shoulder that extends from the position of the excitation laser. We can fit it with an exponential function (Figure S1b). The other background is a broad band between 30 and 80 cm$^{-1}$ (Figure S1a). The exact origin of this feature is unknown. But since it is also present in monolayer (1L) MoS$_2$, it should come from intralayer properties of the material. The broad intralayer band can be fit well with a Gaussian function. The total background can thus be represented by the sum of an exponential function and a Gaussian function. We then subtract the total background from the raw spectrum. Figure 1 shows two examples of the 2L and 5L MoS$_2$ spectra to illustrate our procedure to subtract the Raman background.

We note that the cross-polarized spectra exhibit much weaker relative background signals than the co-polarized spectra (Figure S1e-f). Moreover, the breathing modes are suppressed in the cross-polarized measurement geometry. Therefore, the cross-polarized spectra are effective to reveal the weak shear modes. In the main paper, our analysis of the shear modes focuses on the cross-polarized spectra.

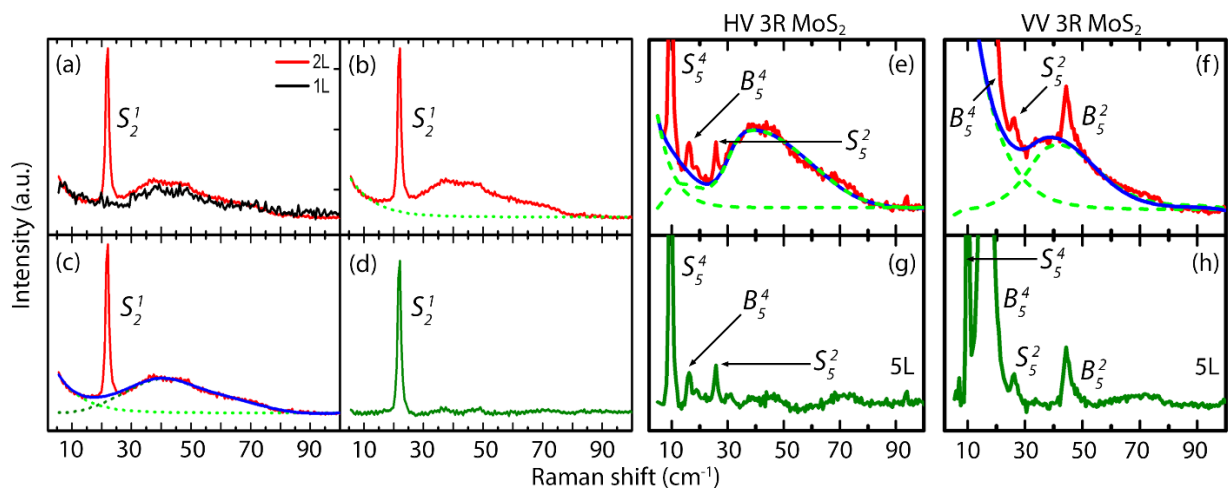

Figure S2. Background subtraction for representative interlayer Raman spectra. (a) Scaled comparison of the cross-polarized Raman spectra of monolayer (1L) MoS$_2$ and bilayer (2L) 3R MoS$_2$. The sharp feature in the 2L spectrum is the interlayer shear mode. The broad band between 30 and 80 cm$^{-1}$, present in both the 1L and 2L spectra, should arise from intralayer properties. (b) The exponential background (dashed line). (c) The total background (blue line), including the broad intralayer band. (d) The 2L spectrum after subtracting the total background. (e, f) The baselines for the cross-polarized (HV) and parallel-polarized (VV) spectra of 5L 3R MoS$_2$, as in panel c. (g-h) The corresponding spectra in panels e and f after subtracting the baselines.

## 2. Individual Raman Spectra for 3R and 2H MoS$_2$

Figures S2 – S5 show the individual Raman spectra for 3R and 2H MoS$_2$ in each layer number in both the parallel-polarized (VV) and cross-polarized (HV) measurement geometries. They correspond to the data in Figure 3 of the main paper.

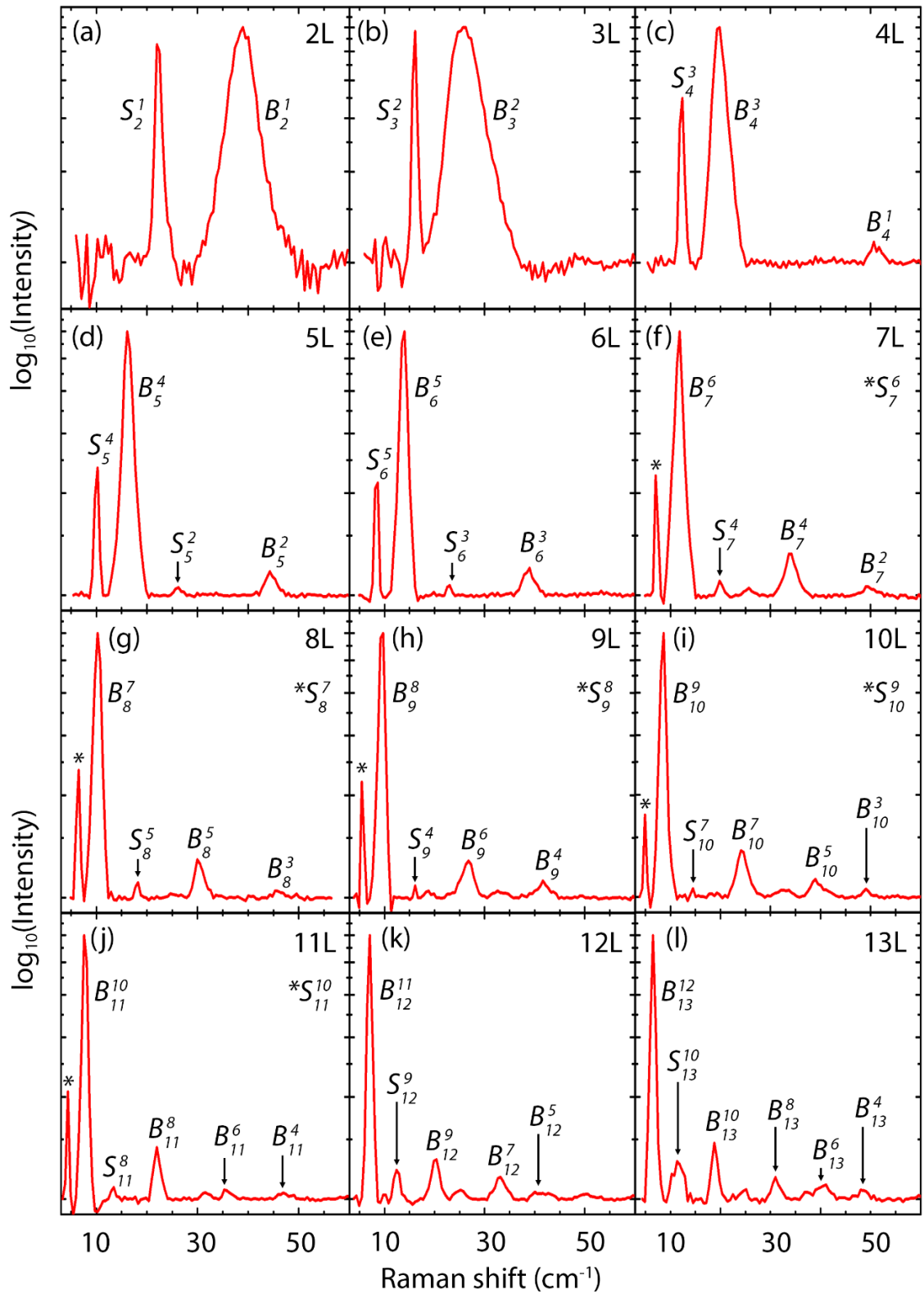

Figure S2. The logarithmic parallel-polarized (VV) Raman spectra of 3R MoS$_2$ from 2L to 13L. The experimental cut-off frequency is ~5 cm$^{-1}$. The lowest-frequency shear modes in the 12L and 13L spectra are cut and unobserved.

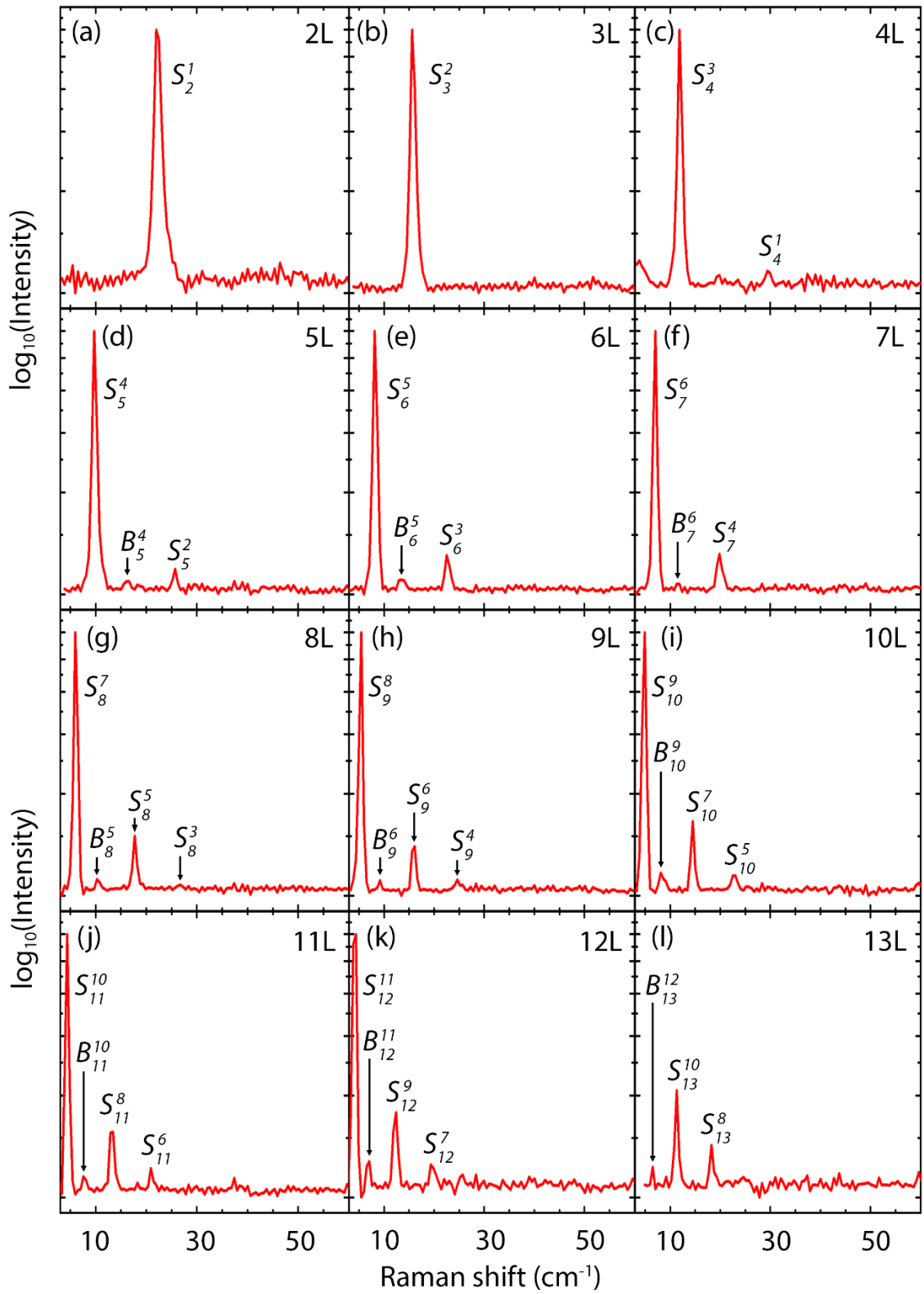

Figure S3. The logarithmic cross-polarized (HV) Raman spectra of 3R MoS$_2$ from 2L to 13L. The experimental cut-off frequency here is ~5 cm$^{-1}$. The lowest-frequency shear mode in the 13L spectrum is cut and unobserved.

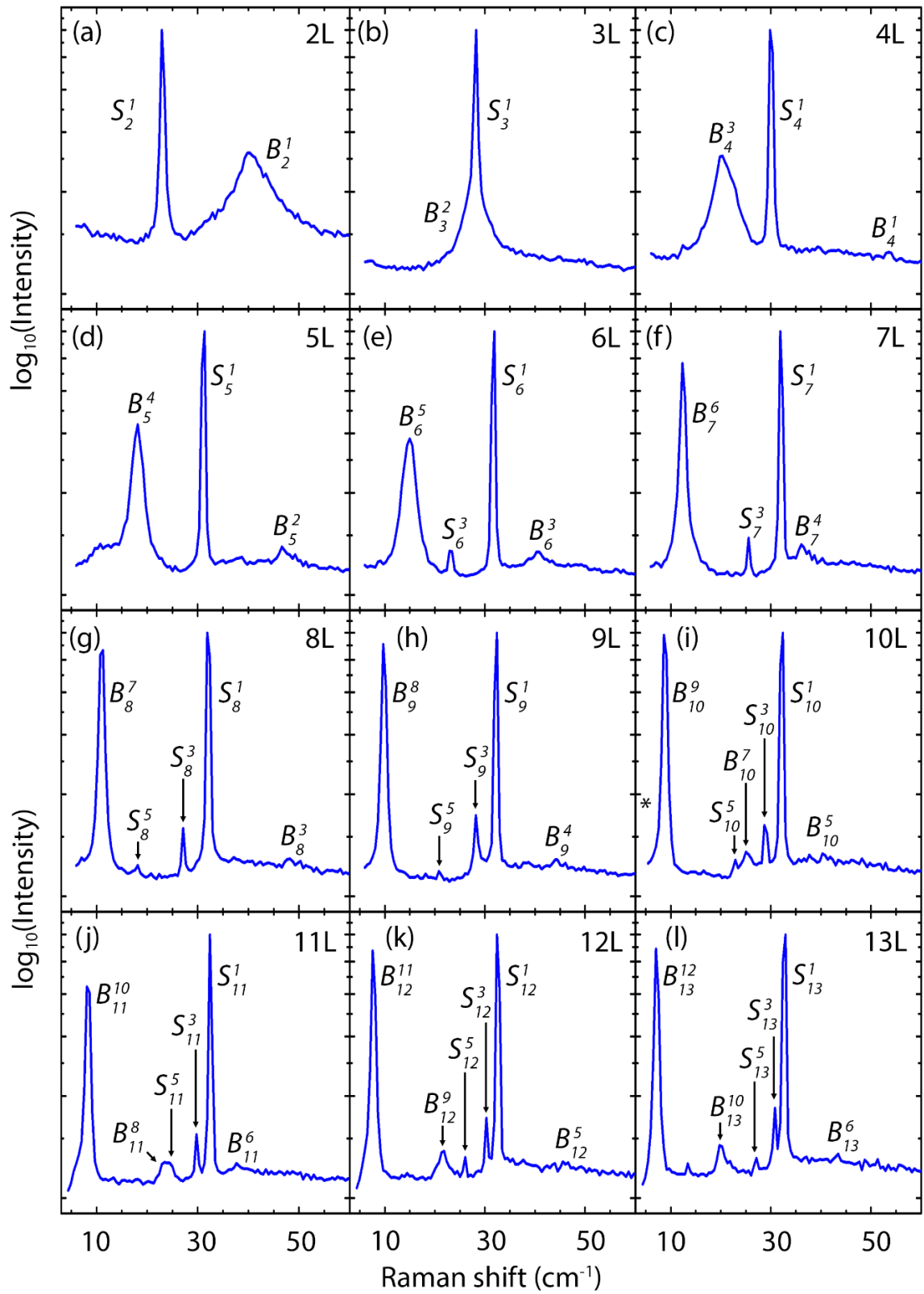

Figure S4. The logarithmic parallel-polarized (VV) Raman spectra of 2H MoS$_2$ with layer thickness from 2L to 13L.

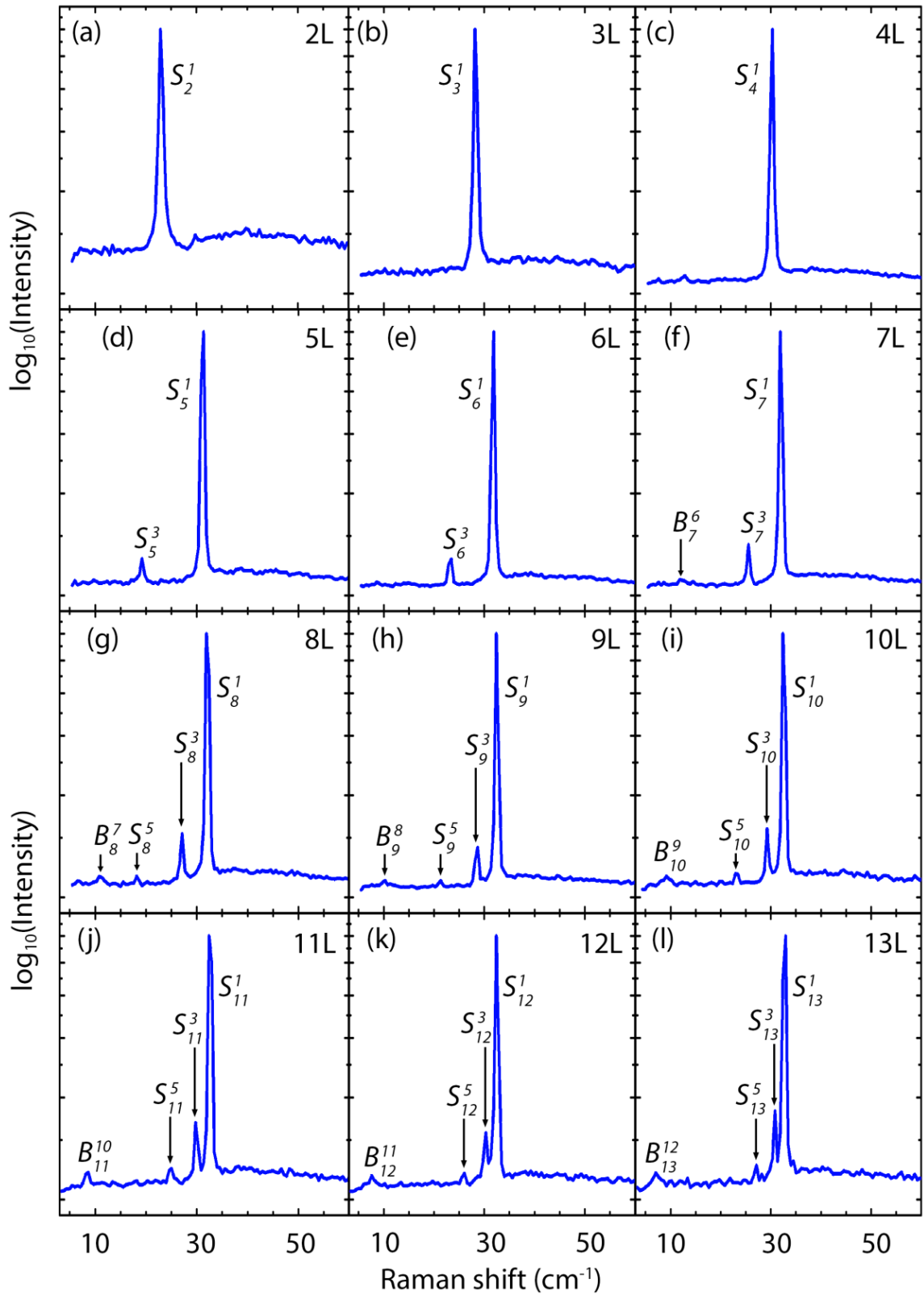

Figure S5. The logarithmic cross-polarized (HV) Raman spectra of 2H MoS$_2$ with layer thickness from 2L to 13L.

## 3. Group-theory analysis of the interlayer phonon modes in 2H and 3R MoS₂

The Raman activity behavior of the interlayer modes in 2H MoS$_2$ (Figure 3 of the main paper) can be explained by the symmetry analysis based on group theory, as reported by prior studies [1, 2]. 2H MoS$_2$ with even layer numbers has the $D_{3d}$ point group with inversion symmetry. The breathing modes $B_N^j$ with branch index $j = 1, 3, 5 \ldots$ have the A$_{1g}$ representation; they are Raman active. $B_N^j$ with $j = 2, 4, 6 \ldots$ have the A$_{2u}$ representation; they are Raman forbidden. 2H MoS$_2$ with odd layer numbers has the $D_{3h}$ point group with plane mirror symmetry. $B_N^j$ with $j = 2, 4, 6 \ldots$ have the $A_1'$ representation; they are Raman active. $B_N^j$ with $j = 1, 3, 5 \ldots$ have the $A_2'$ representation; they are Raman forbidden.

Table S1 displays the Raman activity of the 2H breathing modes. For all layer numbers, the breathing modes are always Raman active in the lowest branch and become alternatively Raman inactive and active for higher branches, as observed in experiment.

|  | 2L | 3L | 4L | 5L | 6L | 7L | 8L | 9L | 10L | 11L | 12L | 13L |
|---|---|---|---|---|---|---|---|---|---|---|---|---|
|  | $D_{3d}$ | $D_{3h}$ | $D_{3d}$ | $D_{3h}$ | $D_{3d}$ | $D_{3h}$ | $D_{3d}$ | $D_{3h}$ | $D_{3d}$ | $D_{3h}$ | $D_{3d}$ | $D_{3h}$ |
| $B_N^1$ | A$_{1g}$, R | A$_1$", IR | A$_{1g}$, R | A$_2$", IR | A$_{1g}$, R | A$_2$", IR | A$_{1g}$, R | A$_2$", IR | A$_{1g}$, R | A$_2$", IR | A$_{1g}$, R | A$_2$", IR |
| $B_N^2$ |  | A$_1'$, R | A$_{2u}$, IR | A$_1'$, R | A$_{2u}$, IR | A$_1'$, R | A$_{2u}$, IR | A$_1'$, R | A$_{2u}$, IR | A$_1'$, R | A$_{2u}$, IR | A$_1'$, R |
| $B_N^3$ |  |  | A$_{1g}$, R | A$_2$", IR | A$_{1g}$, R | A$_2$", IR | A$_{1g}$, R | A$_2$", IR | A$_{1g}$, R | A$_2$", IR | A$_{1g}$, R | A$_2$", IR |
| $B_N^4$ |  |  |  | A$_1'$, R | A$_{2u}$, IR | A$_1'$, R | A$_{2u}$, IR | A$_1'$, R | A$_{2u}$, IR | A$_1'$, R | A$_{2u}$, IR | A$_1'$, R |
| $B_N^5$ |  |  |  |  | A$_{1g}$, R | A$_2$", IR | A$_{1g}$, R | A$_2$", IR | A$_{1g}$, R | A$_2$", IR | A$_{1g}$, R | A$_2$", IR |
| $B_N^6$ |  |  |  |  |  | A$_1'$, R | A$_{2u}$, IR | A$_1'$, R | A$_{2u}$, IR | A$_1'$, R | A$_{2u}$, IR | A$_1'$, R |
| $B_N^7$ |  |  |  |  |  |  | A$_{1g}$, R | A$_2$", IR | A$_{1g}$, R | A$_2$", IR | A$_{1g}$, R | A$_2$", IR |
| $B_N^8$ |  |  |  |  |  |  |  | A$_1'$, R | A$_{2u}$, IR | A$_1'$, R | A$_{2u}$, IR | A$_1'$, R |
| $B_N^9$ |  |  |  |  |  |  |  |  | A$_{1g}$, R | A$_2$", IR | A$_{1g}$, R | A$_2$", IR |
| $B_N^{10}$ |  |  |  |  |  |  |  |  |  | A$_1'$, R | A$_{2u}$, IR | A$_1'$, R |
| $B_N^{11}$ |  |  |  |  |  |  |  |  |  |  | A$_{1g}$, R | A$_2$", IR |
| $B_N^{12}$ |  |  |  |  |  |  |  |  |  |  |  | A$_1'$, R |
|  |  |  |  |  |  |  |  |  |  |  |  |  |
| $S_N^1$ | E$_g$, R | E', R, IR | E$_g$, R | E', R, IR | E$_g$, R | E', R, IR | E$_g$, R | E', R, IR | E$_g$, R | E', R, IR | E$_g$, R | E', R, IR |
| $S_N^2$ |  | E", R* | E$_u$, IR | E", R* | E$_u$, IR | E", R* | E$_u$, IR | E", R* | E$_u$, IR | E", R* | E$_u$, IR | E", R* |
| $S_N^3$ |  |  | E$_g$, R | E', R, IR | E$_g$, R | E', R, IR | E$_g$, R | E', R, IR | E$_g$, R | E', R, IR | E$_g$, R | E', R, IR |
| $S_N^4$ |  |  |  | E", R* | E$_u$, IR | E", R* | E$_u$, IR | E", R* | E$_u$, IR | E", R* | E$_u$, IR | E", R* |
| $S_N^5$ |  |  |  |  | E$_g$, R | E', R, IR | E$_g$, R | E', R, IR | E$_g$, R | E', R, IR | E$_g$, R | E', R, IR |
| $S_N^6$ |  |  |  |  |  | E", R* | E$_u$, IR | E", R* | E$_u$, IR | E", R* | E$_u$, IR | E", R* |
| $S_N^7$ |  |  |  |  |  |  | E$_g$, R | E', R, IR | E$_g$, R | E', R, IR | E$_g$, R | E', R, IR |
| $S_N^8$ |  |  |  |  |  |  |  | E", R* | E$_u$, IR | E", R* | E$_u$, IR | E", R* |
| $S_N^9$ |  |  |  |  |  |  |  |  | E$_g$, R | E', R, IR | E$_g$, R | E', R, IR |
| $S_N^{10}$ |  |  |  |  |  |  |  |  |  | E", R* | E$_u$, IR | E", R* |
| $S_N^{11}$ |  |  |  |  |  |  |  |  |  |  | E$_g$, R | E', R, IR |
| $S_N^{12}$ |  |  |  |  |  |  |  |  |  |  |  | E", R* |

Table S1. The point-group representation of the breathing and shear modes in 2H MoS$_2$. The top row shows the layer numbers of 2H MoS$_2$ and the associated point groups. The left column shows the breathing and shear branches from high to low frequency. "R" means Raman active. "IR" means infrared active. "R*" means that, although the mode is Raman active, it is unobservable in our experimental condition with the laser polarization in the crystal plane. The red color denotes the modes that we observed in Figure 3 of the main paper.

The symmetry property is different for the shear modes. For 2H MoS$_2$ with even layer numbers, the shear modes $S_N^j$ with branch index $j$ = 1, 3, 5 … have the $E_g$ representation; they are Raman active. $S_N^j$ with $j$ = 2, 4, 6 … have the $E_u$ representation; they are Raman forbidden. For 2H MoS$_2$ with odd layer number, $S_N^j$ with $j$ = 1, 3, 5 … have the $E'$ representation; they are Raman active. $S_N^j$ with $j$ = 2, 4, 6 … have the $E''$ representation. Although they are Raman active, their Raman tensors have the form of

$$\begin{pmatrix} 0 & 0 & a \\ 0 & 0 & b \\ a & b & 0 \end{pmatrix}$$

This form of Raman tensors gives zero Raman signal for optical polarization in the $x$-$y$ plane (our experimental conditions). So, these shear modes are unobservable in our experiment. Consequently, the shear modes are always Raman active in the highest branch and become alternatively unobservable and observable for the lower branches in our experimental conditions (Table S1).

The situation is different for 3R MoS$_2$ with lower symmetry. 3R MoS$_2$ has the $C_{3v}$ point group for all the layer numbers, with no inversion symmetry nor mirror symmetry. All breathing modes have the $A_1$ representation and all shear modes have the $E$ representation; they are all Raman active. Therefore, the point-group symmetry alone cannot explain the alternative appearance of the interlayer modes in 3R MoS$_2$ as observed in our experiment.

## 4. Qualitative analysis of Raman activity by inspecting the atomic configurations

In Figure 5 of the main paper, we have analyzed the Raman properties of shear modes in 3L MoS$_2$ by examining the atomic configurations. Here we will elaborate the analysis in the case of 5L MoS$_2$. The same analysis can be readily extended to any layer numbers.

Figure S6a displays the atomic configurations of the four shear modes ($S_5^1$, $S_5^2$, $S_5^3$, $S_5^4$) in 5L MoS$_2$. The 3R $S_5^1$ and $S_5^3$ modes and 2H $S_5^2$ and $S_5^4$ modes have the same configurations for $Q > 0$ and $Q < 0$; they produce zero differential polarizability $\partial \alpha / \partial Q$ and are Raman inactive. But the 3R $S_5^2$ and $S_5^4$ modes and 2H $S_5^1$ and $S_5^3$ modes have different configurations for $Q > 0$ and $Q < 0$; they produce non-zero $\partial \alpha / \partial Q$ and finite Raman response.

We can further compare the two Raman active modes $S_5^2$ and $S_5^4$ in the 3R structure. The lowest-frequency 3R $S_5^4$ mode exhibits rather different atomic configurations between $Q > 0$ and $Q < 0$; it is thus expected to induce a large $\partial \alpha / \partial Q$ and strong Raman response. By contrast, although the higher-frequency 3R $S_5^2$ mode exhibits different configurations for $Q > 0$ and $Q < 0$, the difference is far less obvious than that of the $S_5^4$ mode; it is thus expected to produce a smaller $\partial \alpha / \partial Q$ and weaker Raman response, as observed in our data (Figure S6b). On the other hand, the two Raman-active modes $S_5^1$ and $S_5^3$ in the 2H structure show opposite behavior. The highest-frequency 2H $S_5^1$ mode exhibits more contrasting configurations between $Q > 0$ and $Q < 0$ than the lower-frequency 2H $S_5^3$ mode. The 2H $S_5^1$ mode is therefore expected to be stronger than the $S_5^3$ mode in the Raman spectra, as observed in experiment (Figure S6b).

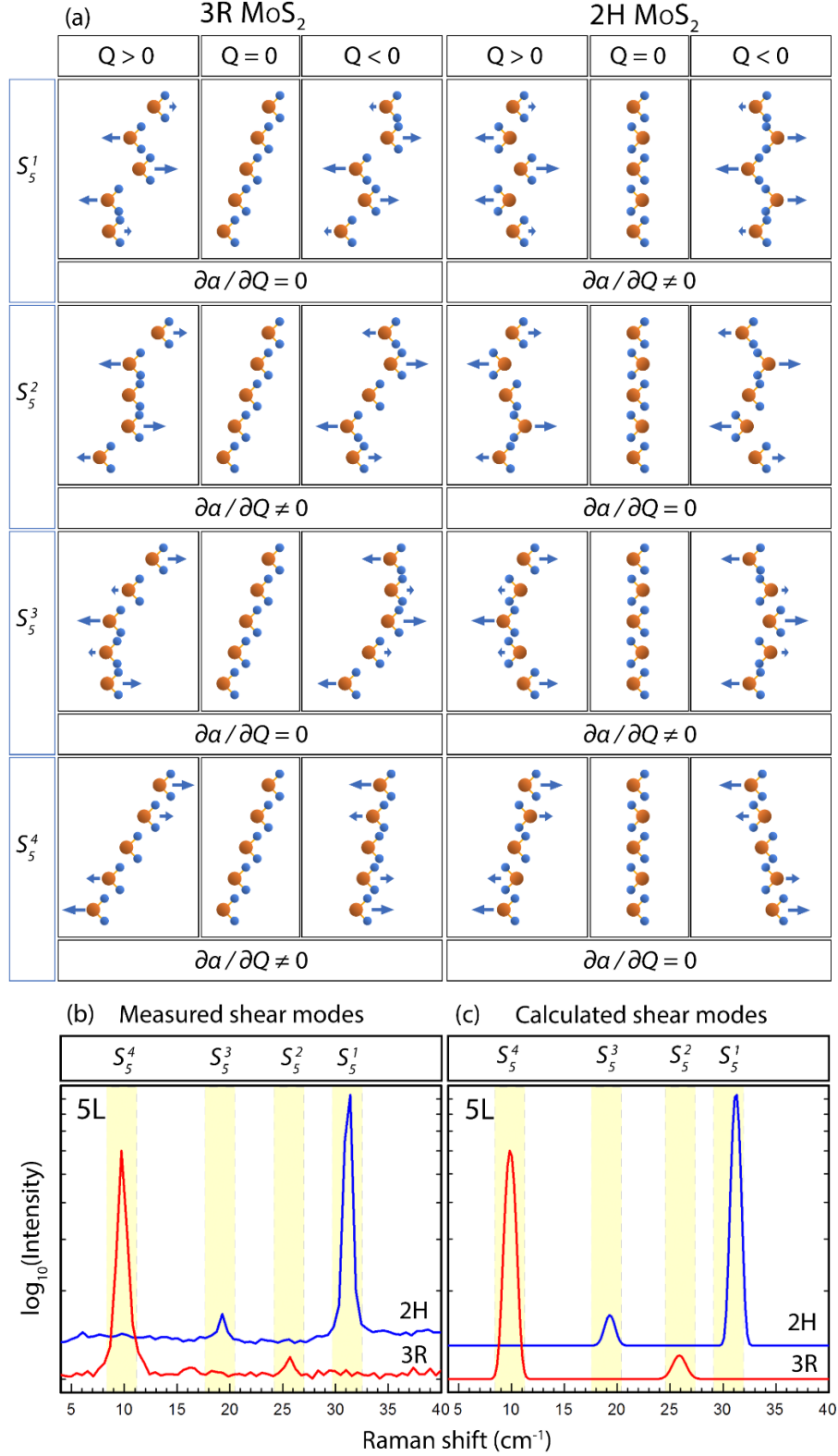

**Figure S6.** (a) Interlayer atomic configurations of shear modes in 5L 3R and 2H MoS$_2$. By comparing the configurations at $Q > 0$ and $Q < 0$, we can deduce whether the differential polarizability $\partial\alpha/\partial Q$ and the Raman response are zero or finite. (b-c) The experimental and simulated cross-polarized Raman shear modes for 5L MoS$_2$.

## 5. The effective bond polarizability model

### 5.1. The general formulation

Our qualitative analysis based on the atomic configurations (Figure S6) can be quantified in an effective bond polarizability model. The detailed derivation of a general bond polarizability model can be found in previous works [3, 4]. Here we summarize the model as it applies to our systems of interest. For a vibration mode $k$, its Raman intensity ($I_k$) is proportional to:

$$I_k \propto \frac{(n_k+1)}{\omega_k} |\hat{e}_i \cdot \tilde{R}(k) \cdot \hat{e}_s^T|^2 \quad \text{(S1)}$$

Here $\omega_k$ is the phonon frequency; $n_k$ is the phonon occupation according to the Bose-Einstein distribution; $\hat{e}_i$ and $\hat{e}_s$ are the polarization unit vectors of the incident and scattered light, respectively; $\tilde{R}(k)$ is the Raman tensor. The diagonal elements of the Raman tensor give rise to parallel-polarized Raman response; the off-diagonal elements of the Raman tensor give rise to cross-polarized Raman response.

In the bond polarizability model, the Raman tensor $\tilde{R}(k)$ is calculated as the sum of contribution from all the bonds distorted by the vibration mode $k$. In the interlayer vibration with rigid layers, the intralayer bonds are unchanged during the vibration; we can thus neglect their contribution to the Raman signal. By considering only the interlayer bonds, the Raman tensor element $R_{\mu\nu}$ ($\mu, \nu = x, y, z$) is calculated as:

$$R_{\mu\nu}(k) = \sum_{l=1}^{N-1} \alpha'_l (x_l - x_{l+1}) \quad \text{(S2)}$$

Here $l$ is the layer index; $x_l$ is the lateral displacement of layer $l$ from its equilibrium position; $\alpha'_l = \partial \alpha_{l,\mu\nu}/\partial x_l$ is the derivative of polarizability of the interlayer bond between layers $l$ and $l+1$ with respect to the $x_l$ displacement. For an interlayer shear mode $S_N^j$ with branch index $j$ in an $N$-layer structure, the relative displacement of each layer is [5]:

$$x_l \propto \cos\left[\frac{(N-j)(2l-1)}{2N}\pi\right] \quad \text{(S3)}$$

The layer displacement is the same for both the 3R and 2H structure. By combining Eq. S1, S2 and S3, we can obtain the relative Raman intensity of the interlayer modes. Below we will consider the cases of 3R and 2H structure.

### 5.2. The bond polarizability model for 3R shear modes

For 3R MoS$_2$, every two adjacent layers have identical interlayer bonds and hence the same $\alpha'_l$ (Figure 6a in the main paper). We may thus remove the layer index $l$ and name it as $\alpha'_l = \alpha'_R$. By Eq. S2, the 3R Raman tensor element is:

$$R_{\mu\nu}(k) = \alpha'_R[(x_1 - x_2) + (x_2 - x_3) + \cdots + (x_{N-1} - x_N)] = \alpha'_R(x_1 - x_N) \quad \text{(S4)}$$

It depends only on the displacement of the top and bottom layers. By putting Eq. S3 into Eq. S4, we obtain:

$$R_{\mu\nu}(k) = \alpha'_R \left\{\cos\left[\frac{(N-j)}{2N}\pi\right] - \cos\left[\frac{(N-j)(2N-1)}{2N}\pi\right]\right\}$$

$$= 2\alpha'_R \sin\left[\frac{(N-j)(N-1)}{2N}\pi\right] \sin\left(\frac{N-j}{2}\pi\right)$$
(S5)

By putting Eq. S5 into Eq. S1, we obtain the Raman intensity of the 3R shear mode $S_N^j$:

$$I_{3R}(k) \propto 4\frac{n_k+1}{\omega_k}|\alpha'_R|^2 \sin^2\left[\frac{(N-j)(N-1)}{2N}\pi\right] \sin^2\left(\frac{N-j}{2}\pi\right)$$
(S6)

By using the identity $\sin^2\left(\frac{N-j}{2}\pi\right) = \frac{1}{2}[1-(-1)^{N-j}]$ for an integer $N-j$, we obtain:

$$I_{3R}(k) \propto 2\frac{n_k+1}{\omega_k}|\alpha'_R|^2 \sin^2\left[\frac{(N-j)(N-1)}{2N}\pi\right] [1-(-1)^{N-j}] \qquad (S7)$$

Eq. S7 corresponds to Eq. 9 in the main paper (with the factor 2 removed). The factor $1-(-1)^{N-j}$ is zero when $N-j$ is an even number. The 3R shear mode is non-zero only for branch indices $j = N-1, N-3, N-5, \ldots$. Such branch-dependent Raman activity matches our observation in experiment (Figure 3, 4 in the main paper).

### 5.3. The bond polarizability model for 2H shear modes

2H MoS$_2$ has a different interlayer bonding configuration from that of 3R MoS$_2$, resulting in different shear Raman modes. As shown in Figure 6b in the main paper, the interlayer bonds in 2H MoS$_2$ exhibit opposite lateral orientation for adjacent layer pairs. This leads to opposite signs of $\alpha'_l$ for adjacent interlayer bonds. By setting $\alpha'_{l=1} = \alpha'_H$, other $\alpha'_l$ would be $(-1)^{l+1}\alpha'_H$. By Eq. S2, the Raman tensor element of a 2H shear mode is:

$$R_{\mu\nu}(k) = \sum_{l=1}^{N-1}(-1)^{l+1}\alpha'_H (x_l - x_{l+1})$$
$$= \alpha'_H[(x_1-x_2)-(x_2-x_3)+(x_3-x_4)-\cdots(x_{N-1}-x_N)]. \qquad (S8)$$

By putting Eq. S3 into Eq. S8, we obtain:

$$R_{\mu\nu}(k) = \alpha'_H \sum_{l=1}^{N-1}(-1)^{l+1}\left\{\cos\left[\frac{(N-j)(2l-1)}{2N}\pi\right] - \cos\left[\frac{(N-j)(2l+1)}{2N}\pi\right]\right\} \qquad (S9)$$

$$= 2\alpha'_H \sin\left(\frac{N-j}{2N}\pi\right) \sum_{l=1}^{N-1}(-1)^{l+1}\sin\left[\frac{(N-j)}{N}l\pi\right] \qquad (S10)$$

$$= -2\alpha'_H \sin\left(\frac{N-j}{2N}\pi\right) \mathrm{Im}\left\{\sum_{l=1}^{N-1}\left[-e^{i\frac{(N-j)}{N}\pi}\right]^l\right\} \qquad (S11)$$

$$= \alpha'_H \sin\left(\frac{N-j}{2N}\pi\right)\frac{(-1)^N \sin\left[\frac{(N-j)(N-1)}{N}\pi\right]+\sin\left(\frac{N-j}{N}\pi\right)}{1+\cos\left(\frac{N-j}{N}\pi\right)} \qquad (S12)$$

We have used:  $\cos A - \cos B = -2\sin\frac{A+B}{2}\sin\frac{A-B}{2}$  (for Eq. S9 → Eq. S10)

$\sin\theta = \mathrm{Im}\{e^{i\theta}\}$  (for Eq. S10 → Eq. S11)

$\sum_{l=1}^{N-1}\gamma^l = \frac{\gamma^N-\gamma}{\gamma-1}$ with $\gamma = -e^{i\frac{(N-j)}{N}\pi}$

$$sin[\pi(N-j)] = 0 \qquad \text{(for Eq. S11} \rightarrow \text{Eq. S12)}$$

Eq. 12 can be simplified by considering:

$$\sin\left[\frac{(N-j)(N-1)}{N}\pi\right] = \sin\left[(N-j)\pi - \frac{N-j}{N}\pi\right]$$
$$= \sin[\pi(N-j)]\cos\left(\frac{N-j}{N}\pi\right) - \cos[\pi(N-j)]\sin\left(\frac{N-j}{N}\pi\right)$$
$$= (-1)^{N-j+1}\sin\left(\frac{N-j}{N}\pi\right) \tag{S13}$$

Here we have used: $\sin(A-B) = \sin A \cos B - \cos A \sin B$
$$\cos[\pi(N-j)] = (-1)^{N-j}$$
$$\sin[\pi(N-j)] = 0$$

By putting Eq. S13 into Eq. S12, we obtain:

$$R_{\mu\nu}(k) = \alpha'_H \sin\left(\frac{N-j}{2N}\pi\right)\frac{\sin\left(\frac{N-j}{N}\pi\right)}{1+\cos\left(\frac{N-j}{N}\pi\right)}[1-(-1)^j] \tag{S14}$$

By using the identity $\frac{\sin A}{1+\cos A} = \tan\frac{A}{2}$, we obtain:

$$R_{\mu\nu}(k) = \alpha'_H \sin\left(\frac{N-j}{2N}\pi\right)\tan\left(\frac{N-j}{2N}\pi\right)[1-(-1)^j] \tag{S15}$$

By putting Eq. S15 into Eq. S1, we obtain the Raman intensity of 2H shear modes:

$$I_{2H}(k) \propto 2\frac{n_k+1}{\omega_k}|\alpha'_H|^2\sin^2\left(\frac{N-j}{2N}\pi\right)\tan^2\left(\frac{N-j}{2N}\pi\right)[1-(-1)^j] \tag{S16}$$

Eq. S16 corresponds to Eq. 10 in the main paper (with the factor 2 removed). The factor $1-(-1)^j$ is zero when the branch index $j$ is an even number; the 2H shear mode is non-zero only for $j = 1$, 3, 5, …. Such branch-dependent Raman activity matches our observation in experiment (Figure 3, 4 in the main paper).

### *5.4. The bond polarizability model for 3R and 2H breathing modes*

The breathing modes of MoS$_2$ are all non-degenerate with one-dimensional group representations. Their Raman tensors are diagonal. Therefore, we only need to consider the diagonal elements of the Raman tensors:

$$R_{xx}(k) = \sum_{l=1}^{N-1} \alpha'_l (z_l - z_{l+1}) \tag{S17}$$

Here $l$ is the layer index; $z_l$ is the vertical displacement of layer $l$ from its equilibrium position. We define $\alpha'_l = \partial\alpha_{l,xx}/\partial z_l$ as the derivative of polarizability of the interlayer bond between layers $l$ and $l+1$ with respect to the $z_l$ displacement. As the vertical differential polarizability does not depend on the lateral orientation of interlayer bonds, $\alpha'_l$ is the same for all layer indices $l$ in either stacking order. We may thus remove the $l$ index and name $\alpha'_l$ as $\alpha'_B$. For a breathing mode $B_N^j$ with branch index $j$ in an $N$-layer structure, the normal-mode layer displacement is:

$$z_l \propto \cos\left[\frac{(N-j)(2l-1)}{2N}\pi\right] \tag{S18}$$

The differential polarizability and layer displacement of breathing modes follow the same forms as those of 3R shear modes (Eq. S3, S4). Therefore, the Raman intensity of both 2H and 3R breathing modes must show the same expression as that of 3R shear modes (Eq. S7):

$$I_{3R}(k) \propto 2\frac{n_k+1}{\omega_k}|\alpha'_B|^2 \sin^2\left[\frac{(N-j)(N-1)}{2N}\pi\right]\left[1-(-1)^{N-j}\right] \tag{S19}$$

Here $\alpha'_B$ corresponds to the vertical differential polarizability of each interlayer bond, in either 2H or 3R stacking order. The factor $1-(-1)^{N-j}$ is zero when $N-j$ is an even number; the breathing mode is non-zero only for branch indices $j = N-1, N-3, N-5 \ldots$ for both the 2H and 3R structure. Such branch-dependent Raman activity matches our observation in experiment (Figure 3, 4 in the main paper).

### 5.5. Example of 5L MoS$_2$ shear modes

We can illustrate our analysis and modelling by using the example of 5L MoS$_2$, which have four shear modes with branch index $j = 1, 2, 3, 4$. For the shear mode $S_{N=5}^j$, the displacement of each layer ($l = 1 - 5$) is:

$$x_l = \cos\left[\frac{(5-j)(2l-1)}{10}\pi\right]. \tag{S20}$$

Figure S6a shows the schematic layer displacements for the 5L 3R and 2H structures. By using Eq. S7 and S16, we can calculate the relative intensity of the shear modes in 5L MoS$_2$ for both 3R and 2H stacking order. The results are plotted in Figure S6c.

### 5.6. Comparison between theoretical and experimental Raman intensity

By using the effective bond polarizability model, we can calculate the relative Raman intensity of the shear and breathing modes in 2H and 3R MoS$_2$ for different layer numbers. Our theory quantitatively reproduces the stacking-order and branch-index dependence of the interlayer-mode Raman intensity observed in our experiment (Figures S7, S8).

Figure S7 compares the theoretical and experimental Raman intensities of the Raman-active shear branches in 2H and 3R MoS$_2$ for layer number $N = 4 - 13$L. All intensities are normalized with respect to the strongest shear branch. The experimental data are the integrated area of Gaussian fits for the measured shear modes in Figure S3 and S5. They exhibit two remarkable features. First, the Raman intensity decreases almost exponentially for successive Raman-active shear branches. Second, the 2H and 3R structures exhibit opposite branch dependence – the intensity decreases with increasing (decreasing) branch index in 2H (3R) stacking. Our theory reproduces quantitatively both features in all layer numbers.

Similarly, Figure S8 compares the normalized theoretical and experimental Raman intensities of breathing modes in 2H and 3R MoS$_2$. The experimental data are the integrated areas of Gaussian fits for the measured breathing modes in Figure S2 and S4. Unlike the shear modes, the breathing modes exhibit the same branch dependence in 2H and 3R $MoS_2$ – both intensities show almost exponential decrease with decreasing branch index. Our theory agrees well with experiment for thickness up to 9 layers. For thickness higher than 9 layers, the experimental intensity starts to deviate from the theoretical values for the weakest breathing branches. We are uncertain about such deviation. Further investigations are needed to resolve and understand the weak breathing modes.

For 4L and 5L $MoS_2$, we have also compared our results with the predicted Raman intensity from density-functional theory (DFT) (triangles in Figures S7, S8; see Section S6 for more discussion of the DFT calculation). The results of the bond polarizability model match the experimental data slightly better than the DFT calculation.

## 5.7. *Simulation of parallel-polarized interlayer Raman spectra*

Figures S9 and S10 represent our simulation of the parallel-polarized (VV) interlayer Raman spectra, in comparison with experiment, for both 3R and 2H $MoS_2$ with layer numbers 3L, 5L, 7L, 9L, 11L, and 13L (the simulation for even layer numbers is similar). In the simulation, we combine the frequencies calculated by the linear chain model (Figure 4) and the relative Raman intensity calculated by the bond polarizability model (Figures S7 and S8). The intensity ratio between the whole set of shear modes and the whole set of breathing modes are adjusted according to experiment.

As our simple theory cannot predict the Raman line width, we adopt the experimental line width directly, which are extracted by Gaussian fits. To compare our theory with experiment, we plot the calculated spectra by a sum of Gaussian functions with the same width as the experimental spectra. For those calculated modes that are not observed in experiment, we assign a line width based on the width of similar features in experiment. As shown in Figure S9 and S10, our simulation reproduces most of the Raman features observed in experiment.

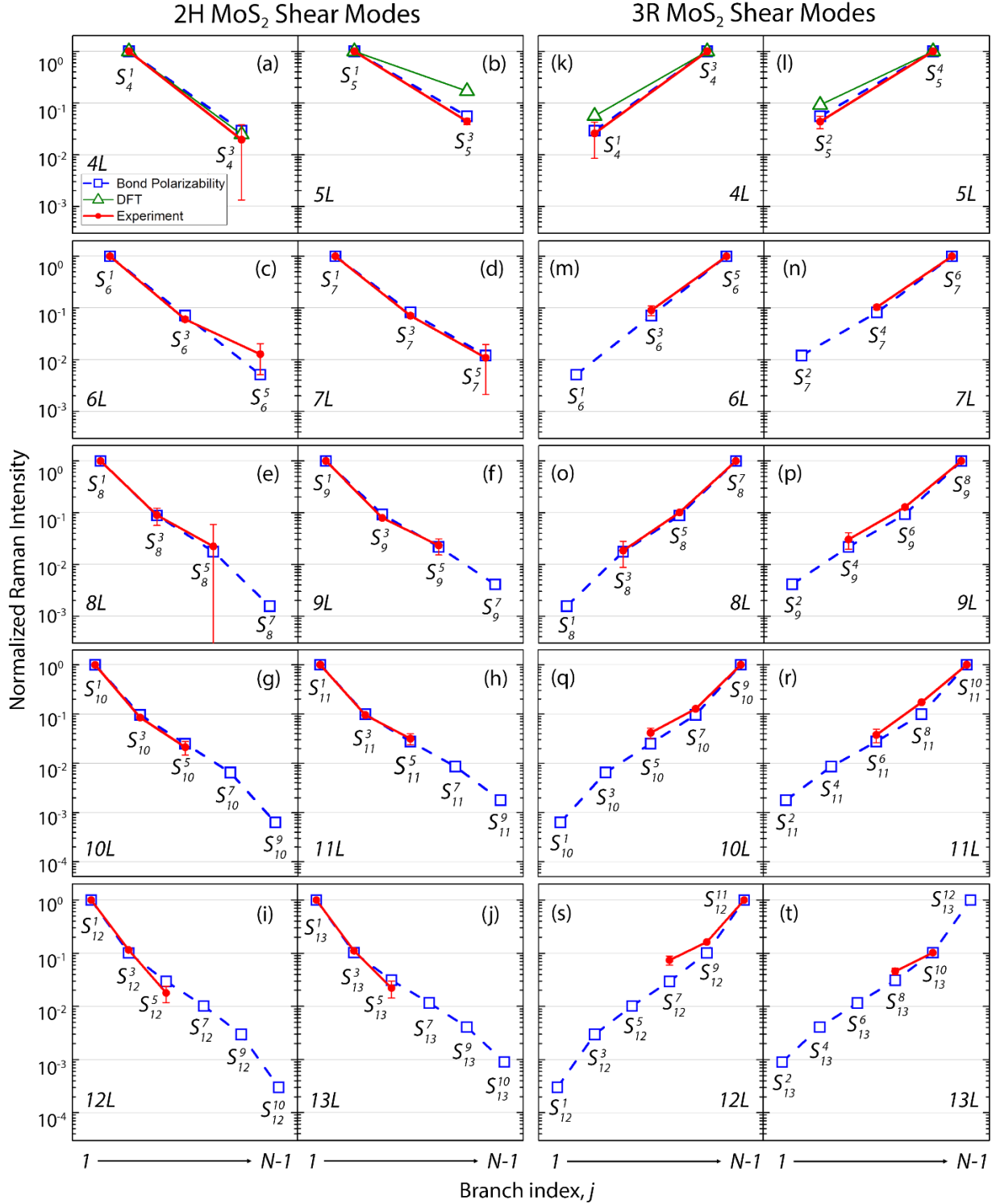

Figure S7. The relative Raman intensity of successive Raman-active shear branches in (a-j) 2H MoS$_2$ and (k-t) 3R MoS$_2$. The branch index $j$ = 1, 2, 3, … $N$-1 counts from high to low frequency. The solid dots are the experimental integrated intensity of the Raman peaks in Figures S3 and S5. The open squares are the theoretical Raman intensity of the effective bond polarizability model (calculated by Eq. S7, S16). The open triangles are the results from first-principles calculations (4L and 5L only). As the $j = N − 1$ shear branch is not observed in the 13L 3R MoS$_2$ sample, we normalize the experimental intensity to the $j = N − 3$ branch.

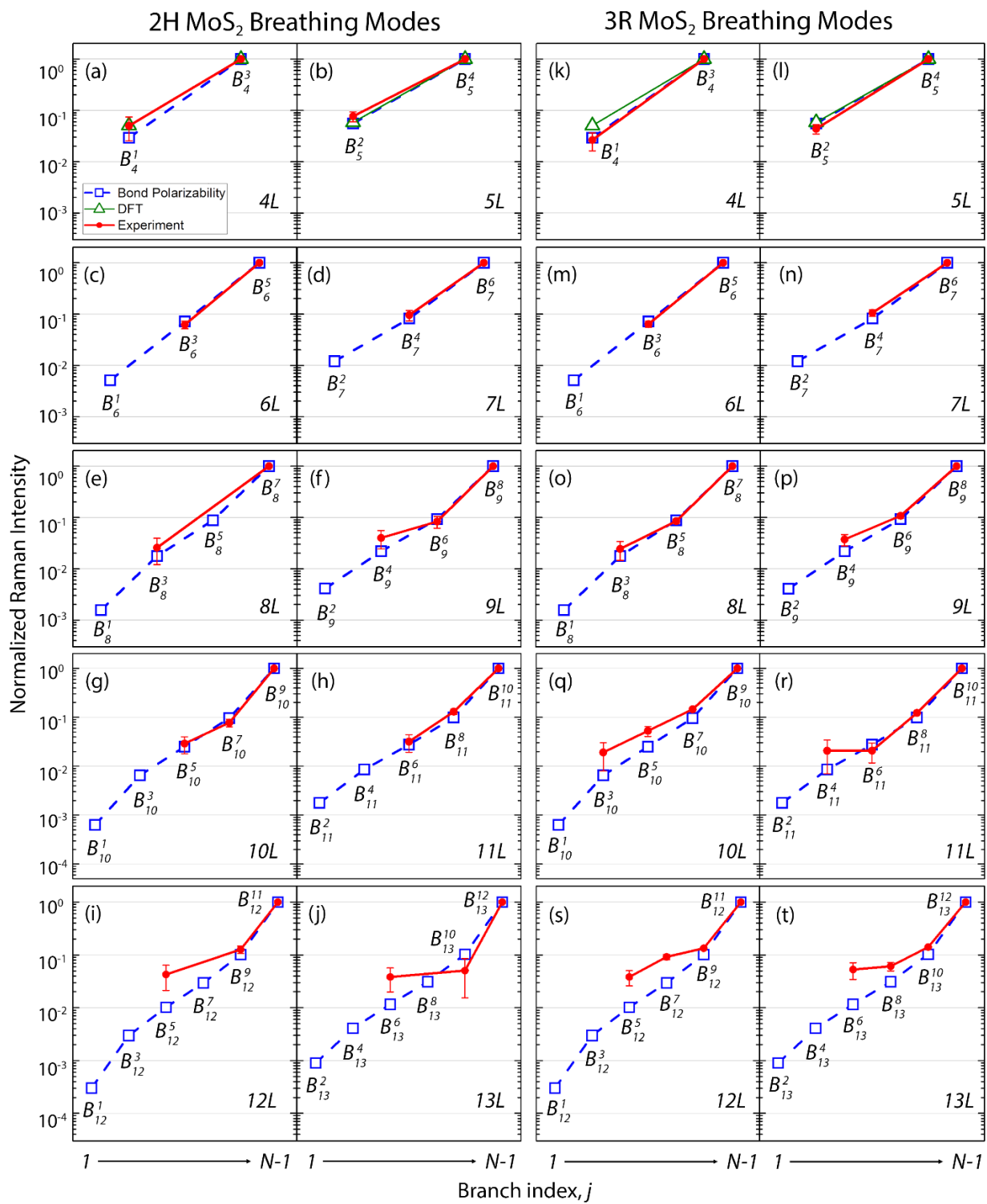

Figure S8. The relative Raman intensity of successive Raman-active breathing branches in (a-j) 2H MoS$_2$ and (k-t) 3R MoS$_2$. The branch index $j = 1, 3, 5, \ldots N$-1 counts from high to low frequency. The solid dots are the experimental integrated intensities of the breathing modes in Figures S2 and S4. The open squares are the calculated Raman intensity by the effective bond polarizability model (Eq. S19). Open triangles are results from the first-principles calculations (4L and 5L only).

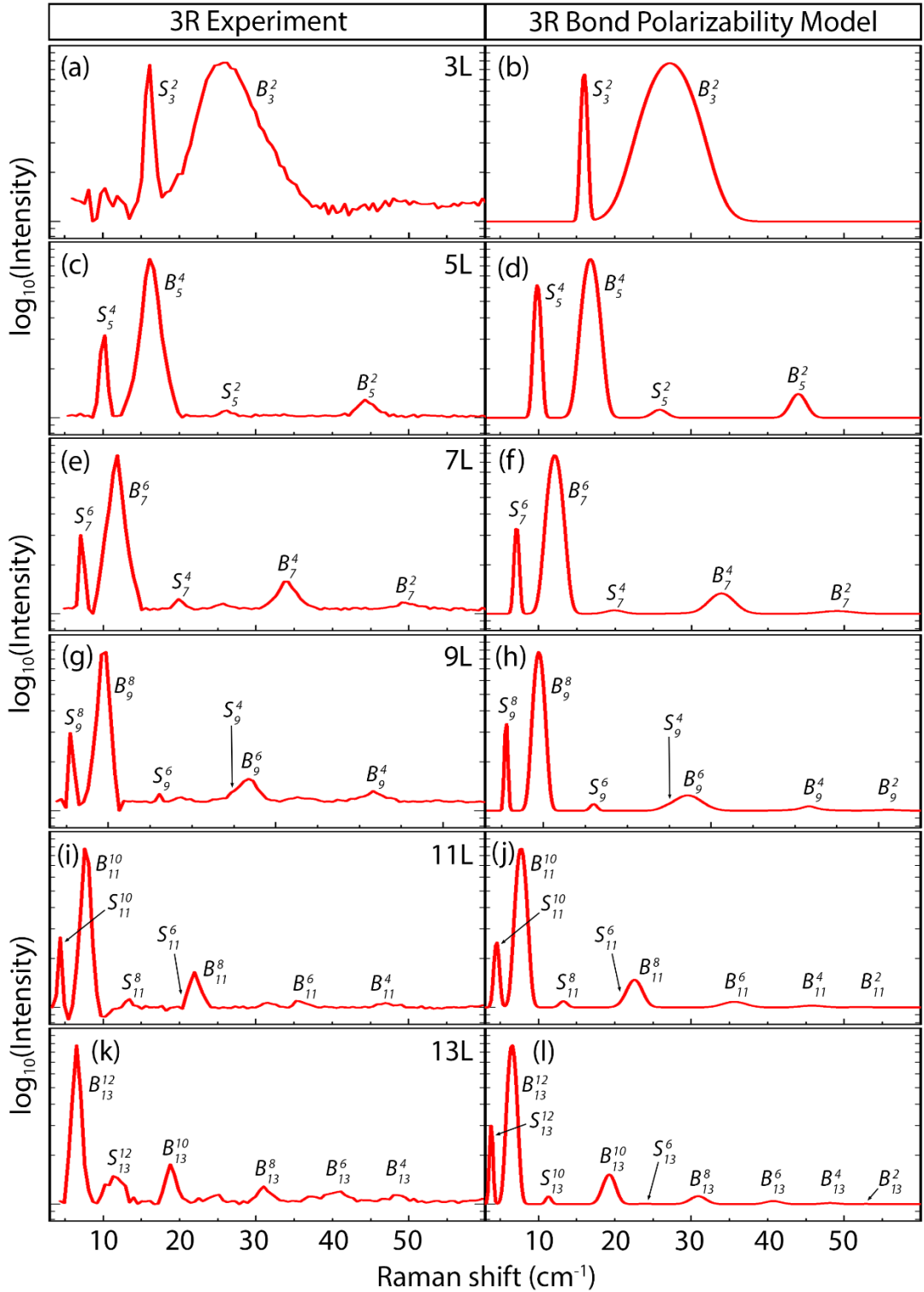

Figure S9. Comparison of experimental and simulated parallel-polarized (VV) Raman spectra for 3R MoS$_2$ with layer number (a,b) 3L, (c,d) 5L, (e,f) 7L, (g,h) 9L, (i,j) 11L, and (k,l) 13L. The theoretical spectra are plotted as a sum of Gaussian functions. The theoretical Raman intensity and frequency are calculated from the bond polarizability model and the linear chain model, respectively. The line widths of the theoretical peaks are made to match the experimental line widths.

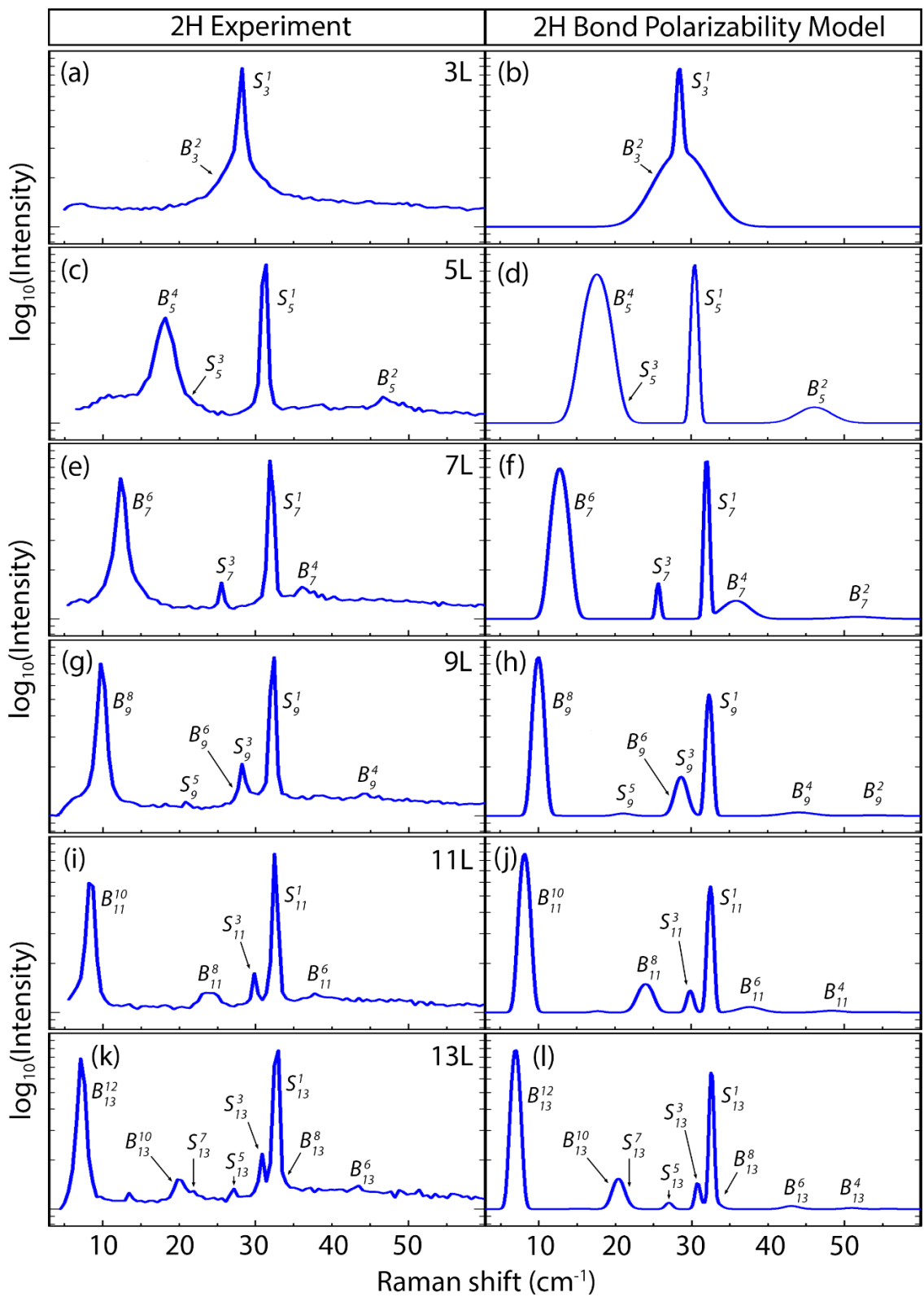

Figure S10. Comparison of experimental and simulated parallel-polarized (VV) Raman spectra for 2H MoS$_2$ as in Figure S9.

## 6. The first-principles calculations of interlayer Raman modes for 2H and 3R MoS$_2$

Although the bond polarizability model is effective to calculate the relative Raman intensity of different branches of shear (breathing) modes, it is unable to calculate the relative Raman intensity between the shear modes and breathing modes. To this end, we turn to the first-principles calculations. We have carried out density-functional theory (DFT) [6, 7] calculations for 2H and 3R MoS$_2$ from bilayer to five-layer thickness by using the Quantum ESPRESSO (QE) code [8] and VASP [9, 10] with the local-density-approximation exchange-correlation functional. A shifted Monkhorst-Pack uniform $k$-grid of 12×12×1 is employed for all structural optimizations and self-consistent calculations. A vacuum region of more than 20 Å is introduced along the out-of-plane direction to eliminate spurious interactions between periodic images. All the atomic structures and unit cells are fully relaxed until the stress along each axis is smaller than 0.5 kbar and the forces on the atoms are smaller than 0.002 eV/ Å. After the structures are fully relaxed, we calculate the phonon frequencies for 2H and 3R MoS$_2$ at the Brillouin zone center.

The predicted frequencies by QE and VASP are listed in Table S2 and plotted in Figure S11 and S12. By comparison with experiment, the VASP frequencies are found to match the experimental frequencies better than the QE frequencies (Figure S11). In addition, only the VASP calculation can reproduce the fact that the 2H breathing modes are consistently ~3 cm$^{-1}$ higher than the 3R breathing modes (Figure S12c; Figure 4). Therefore, we choose to use the frequencies predicted by VASP. However, QE is more effective to calculate the Raman intensity [11]. We therefore use QE to calculate the Raman tensor and Raman intensity of the interlayer modes.

|  | 2H MoS$_2$ | | | | | | | |
|---|---|---|---|---|---|---|---|---|
|  | **Shear mode frequency (cm$^{-1}$)** | | | | **Breathing mode frequency (cm$^{-1}$)** | | | |
| 2L | 26.8 (25.1) | | | | 41.5 (40.0) | | | |
| 3L | 19.4 (17.5) | 34.8 (30.9) | | | 29.8 (28.2) | 50.2 (48.8) | | |
| 4L | 15.7 (13.3) | 26.8 (25.0) | 34.6 (32.9) | | 23.4 (21.5) | 41.7 (39.8) | 52.7 (51.9) | |
| 5L | 13.6 (11.1) | 23.1 (21.1) | 30.5 (29.1) | 37.8 (34.4) | 16.4 (17.8) | 33.4 (33.6) | 46.1 (46.2) | 54.9 (54.4) |
|  | 3R MoS$_2$ | | | | | | | |
|  | **Shear mode frequency (cm$^{-1}$)** | | | | **Breathing mode frequency (cm$^{-1}$)** | | | |
| 2L | 27.9 (25.8) | | | | 40.1 (37.1) | | | |
| 3L | 20.2 (17.6) | 34.0 (31.6) | | | 28.3 (26.2) | 46.4 (45.5) | | |
| 4L | 18.3 (13.6) | 29.6 (25.7) | 37.4 (33.8) | | 23.8 (20.0) | 38.9 (37.1) | 49.4 (48.6) | |
| 5L | 19.3 (11.6) | 26.3 (21.9) | 32.7 (30.2) | 37.3 (35.7) | 18.6 (16.6) | 34.8 (31.7) | 45.3 (43.6) | 53.5 (51.4) |

Table S2 The calculated frequencies of shear and breathing modes in 2H and 3R MoS$_2$ from 2L to 5L by using QE (the first number) and VASP (the second number in parenthesis).

Figure S13 shows the predicted parallel-polarized (VV) Raman spectra of the interlayer modes in 2H and 3R MoS$_2$ with thickness $N = 2 – 5L$. In the calculation, we have used VASP to obtain the frequency, but QE to obtain the Raman intensity. Our results are comparable with the experimental data as well as the results of the bond polarizability model (triangles in Figures S7,

S8 for 4L and 5L MoS$_2$). Figure 2 in the main paper shows the unpolarized Raman spectra of the interlayer modes for bilayer 3R and 2H MoS$_2$, which are the sum of the parallel-polarized and cross-polarized spectra in our calculation. The calculated 2L MoS$_2$ Raman spectra agree well with the experimental ones.

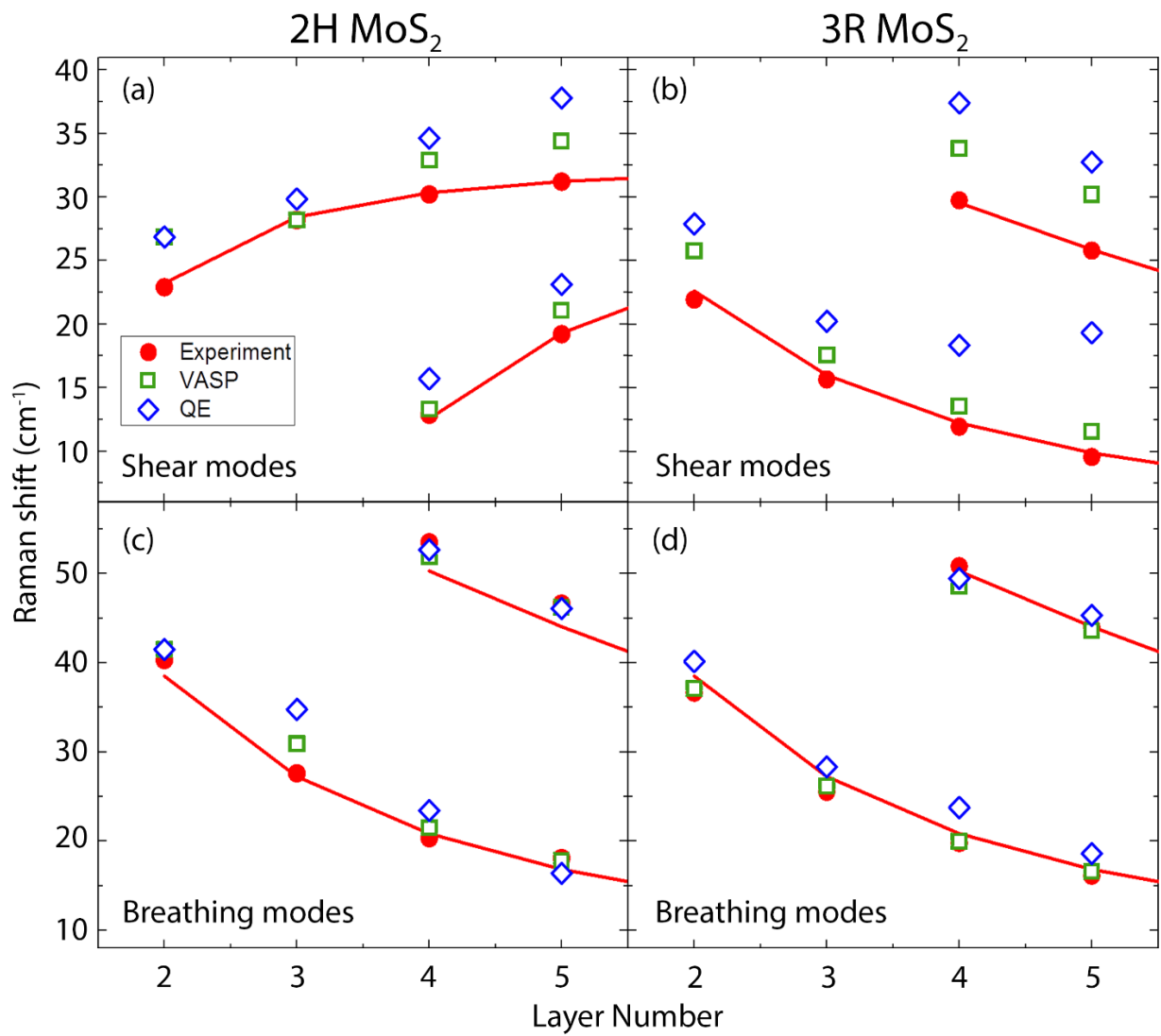

Figure S11. Comparison of DFT and experimental frequencies of (a,b) shear modes and (c,d) breathing modes for (a,c) 2H MoS$_2$ and (b,d) 3R MoS$_2$. The red dots are experimental data. The open green squares and blue diamonds are the predicted frequencies by VASP and QE, respectively. The VASP frequencies generally agree better with the experimental data than the QE frequencies.

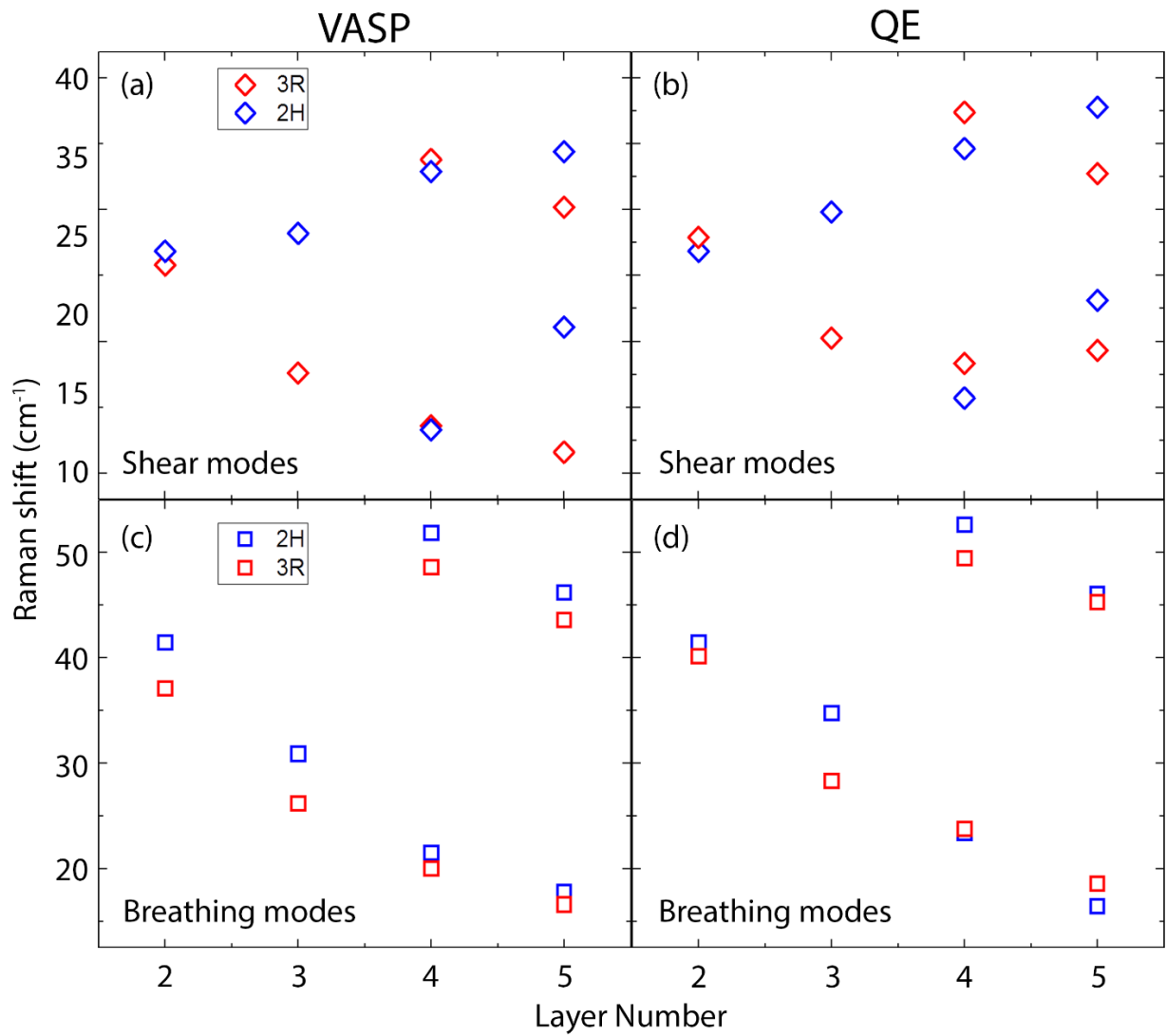

Figure S12. Comparison of the calculated frequencies of (a, b) shear modes and (c, d) breathing modes by using (a, c) VASP and (b, d) QE for 2H and 3R $MoS_2$. Only the VASP calculation reproduces the experimental results that the 2H breathing modes are consistently higher in frequency than the 3R breathing modes (panel c).

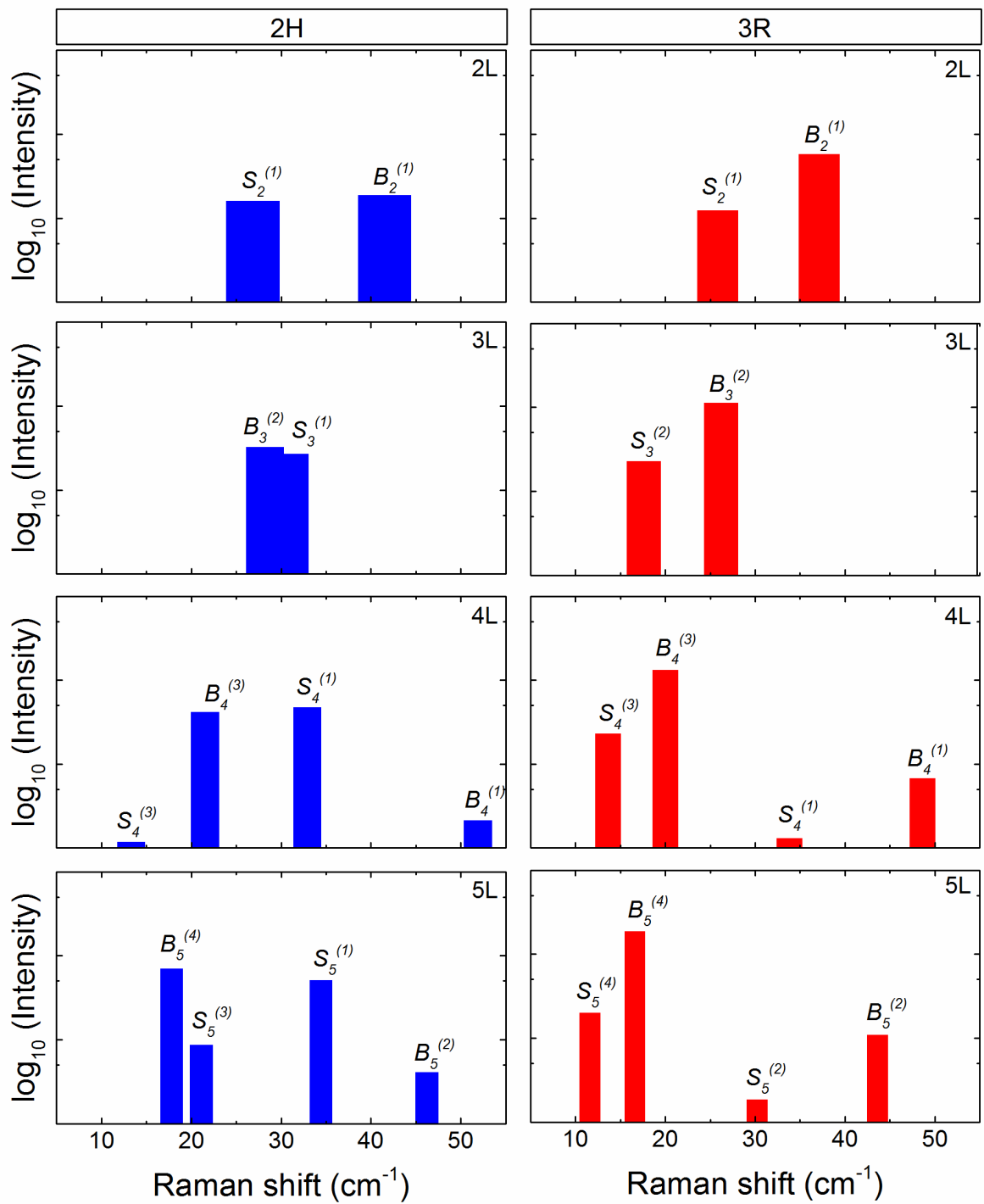

Figure S13. DFT calculated Raman spectra for 2H and 3R MoS$_2$ with parallel polarization.

**Supplementary References**